\documentclass[manuscript]{emulateapj}
\usepackage{lscape}

\shorttitle{BLAST: the Redshift Survey}
\shortauthors{Eales et al.}

\begin{document}


\title{BLAST: the Redshift Survey}


\author{
Stephen Eales\altaffilmark{1},
Edward L. Chapin\altaffilmark{2},
Mark J. Devlin\altaffilmark{3},
Simon Dye\altaffilmark{1},
Mark Halpern\altaffilmark{2},
David H. Hughes\altaffilmark{4},
Gaelen Marsden\altaffilmark{2},
Philip Mauskopf\altaffilmark{1},
Lorenzo Moncelsi\altaffilmark{1},
Calvin B. Netterfield\altaffilmark{6,7},
Enzo Pascale\altaffilmark{1},
Guillaume Patanchon\altaffilmark{8},
Gwenifer Raymond\altaffilmark{1},
Marie Rex\altaffilmark{3},
Douglas Scott\altaffilmark{2},
Christopher Semisch\altaffilmark{3},
Brian Siana\altaffilmark{5}
Matthew D. P. Truch\altaffilmark{3},
\& Marco P. Viero\altaffilmark{6}
}

\altaffiltext{1}{Cardiff University, School of Physics \& Astronomy,
Queens Buildings, The Parade, Cardiff, CF24 3AA, U.K.}
\altaffiltext{2}{Department of Physics \& Astronomy, University of
British Columbia, 6224 Agricultural Road, Vancouver, BC V6T 1Z1, Canada}
\altaffiltext{3}{Department of Physics \& Astronomy, University of
Pennsylvania, 209 South 33rd Street, Philadelphia, PA, 19104, U.S.A.}
\altaffiltext{4}{Instituto Nacional de Astrof\'{i}sica \'{O}ptica
y Electr{\'o}nica (INAOE), Aptdo. Postal 51 y 72000 Puebla, Mexico}
\altaffiltext{5}{California Institute of Technology, MS 105-24, Pasadena,
CA 91125, U.S.A.}
\altaffiltext{6}{Department of Astronomy \& Astrophysics, University of
Toronto, 50 St. George Street Toronto, ON M5S 3H4, Canada}
\altaffiltext{7}{Department of Physics, University of Toronto, 60 St.
George Street, Toronto, ON M5S 1A7, Canada}
\altaffiltext{8}{Universit{\'e} Paris Diderot, Laboratoire APC, 10,
rue Alice Domon et L\'{e}onie Duquet 75205 Paris, France}



\begin{abstract}
The Balloon-borne Large Aperture Submillimeter Telescope (BLAST) has recently
surveyed $\simeq$8.7 deg$^2$ centered on GOODS-South at 250, 350 and 500
$\mu$m. In Dye et al. (2009) we presented the catalogue of
sources detected at $\rm 5\sigma$ in at least one band in this field
and the probable counterparts to these sources in other wavebands. In this
paper, we present the results of a redshift survey in which we
succeeded in measuring redshifts for 82 of these counterparts. 
The spectra show that the BLAST counterparts are mostly star-forming
galaxies but not extreme ones when compared to those
found in the Sloan Digital Sky Survey. 
Roughly one quarter of the BLAST counterparts contain an active nucleus.
We have used
the spectroscopic redshifts to carry out a test
of the ability of photometric redshift methods to estimate the
redshifts of dusty galaxies, showing that the standard methods work well even
when a galaxy contains a large amount of dust.
We have also investigated the cases where there are
two possible counterparts to the BLAST source, finding that in at least half
of these there is evidence that
the two galaxies are physically associated, either
because they are interacting or because 
they are in the same large-scale structure. 
Finally, we have made the first direct measurements of the
luminosity function in the three BLAST bands. We find strong evolution out to
$\rm z=1$, in the sense that there is a large increase in the space-density
of the most luminous galaxies. We have also investigated the evolution of the dust-mass
function, finding similar strong evolution in the space-density of the galaxies with
the largest dust masses, showing that the luminosity evolution
seen in many wavebands is associated with an increase in the reservoir of interstellar
matter in galaxies.

\end{abstract}


\keywords{galaxies: evolution --- surveys --- submillimeter --- galaxies: high-redshift}



\section{Introduction}

Excluding the cosmic microwave background, 
the main peaks in the extragalactic background radiation
are in the optical and far-IR/submillimetre wavebands with
roughly the same amount of 
energy in each \citep{dwek98,fix98}, implying that approximately half
of the total energy emitted by galaxies since their formation has been absorbed by dust and
then reradiated at longer wavelengths. This energy budget strongly suggests that to
completely understand galaxies and their evolution it is crucial
to understand the nature of the sources that make up the cosmic infrared background
(henceforth the CIB). However, in the
13 years since the discovery of this background \citep{puget} it has proved difficult to answer this 
question,
partly because of the technical challenges of working at these wavelengths
and partly because the atmosphere is opaque over much of the wavelength range from
20 $\mu$m to 1 mm, with only the 850-$\mu$m atmospheric window having routine transmission of over 50\%.

After the discovery of the CIB, much of the early progress in determining the nature of the
sources that compose it came from the ground-based surveys with the SCUBA camera
on the James Clerk Maxwell Telescope \citep{sib,h98,amy,eales99}. These surveys resolved about
30\% of the background at 850 $\mu$m and close to 100\% of the background
if one includes the small numbers
of sources detected in lensing surveys \citep{blain99,knudsen}. However, the full potential
of these surveys has been hard to achieve due to the
poor angular resolution combined with the faintness of the optical counterparts,
which has made it a challenge both to identify the correct optical counterparts
and to measure their redshifts.
The most extensive redshift 
survey of the SCUBA surveys \citep{chap2005} found a median redshift
of $\simeq$2.2, and in general the SCUBA sources are luminous dusty galaxies seen in the
early universe that are even more luminous
than the Ultraluminous Infrared Galaxies (ULIRGs) found in the universe today 
\cite{cop2008}.
Evidence from X-ray observations \citep{alex2005} and from mid-infrared spectroscopy 
with Spitzer \citep{pope2008,menend1,menend2} suggests
that while a large fraction of these sources appear to contain active nuclei, most of the energy
emitted by these objects ultimately comes from young stars rather than an obscured active nucleus.
The star-formation rates implied by the luminosities of these objects are often as much
as 1000 $\rm M_{\odot}\ year^{-1}$ \citep{alex2005}, enough to build a large galaxy in only 1\% of the age
of the universe. Many authors have argued that the space-density of these sources and their implied
star-formation
rates show that they are probably the ancestors of present-day elliptical galaxies \citep{lilly,scott02,dunne2003}.

The SCUBA surveys, however, had two major limitations. First, the energy in the background ($\rm I_{\nu} \nu$) at
850 $\mu$m is only one thirtieth of the energy in the background at its peak at $\simeq$200 $\mu$m, and
so the sources detected in the SCUBA surveys may not be representative of the CIB as a whole. 
Dye et al. (2007) used a stacking argument to show that the sources that constitute 30\% of the background
at 850$\mu$m make up at most 18\% of the background at 160 $\mu$m.
Chapman et al. (2005) used indirect arguments to reach the stronger conclusion
that the sources
that make up 60\% of the background at 850 $\mu$m contribute only
6\% of the background at 200 $\mu$m.
Second, since most of the
sources detected in the SCUBA surveys are at very high redshifts, we actually know remarkably
little about the submillimetre properties of the nearby universe. To produce
a fair sample of the nearby universe that is not biased by the presence of
a small number of clusters or unusually empty regions,
it is necessary
to survey a large area of sky, which was not possible with SCUBA because of its small
field of view. Therefore, estimates of the local luminosity function
at submillimetre wavelengths ($\rm 100\ \mu m < \lambda < 1\ mm$), which are crucial for
investigating the cosmic evolution in this waveband,
are based either on extrapolations from the survey with the IRAS satellite at shorter
wavelengths or on
submillimetre observations of samples of galaxies selected in other wavebands \citep{dunne2000,cat2005}.
Both of these approaches have obvious drawbacks.

There has recently been a major step forward in this field as the result of observations
with the Balloon-borne Large Aperture Submillimeter Telescope (BLAST) \citep{devlin}.
BLAST has carried out surveys of the extragalactic sky in two fields, one near
the South Ecliptic Pole and one centered on the southern field of the Great Observatories
Origins Deep Survey (GOODS-South). Each survey covered about 10 deg$^2$, and, for comparison,
the largest SCUBA survey only covered $\simeq$0.33 deg$^2$ \citep{cop2006}. The BLAST surveys
were at three wavelengths---250, 350 and 500 $\mu$m---and since the shortest wavelength
is close to the peak of the CIB, the sources detected in these surveys are likely to
be more representative of the CIB than the sources detected in the SCUBA surveys. 

The BLAST survey of GOODS-South has been particularly useful because of the
wealth of data at other wavelengths that exists in this field. 
There have been several studies of the
statistical properties in the BLAST bands of sources from
catalogues defined from Spitzer 24-$\mu$m observations \citep{devlin,marsden,enzo}. These 
have shown that the 24-$\mu$m sources may well contribute all of the
CIB (see also Dole et al. 2006). Therefore, whereas the sources found in samples at one end (850$\mu$m)
of the far-IR/submillimetre waveband are not representative of the CIB, those
at the other end (24 $\mu$m) do seem to be.
By combining the BLAST results with Spitzer 70-$\mu$m data and a mixture of
photometric and spectroscopic redshifts, Pascale et al. (2009) made the
first direct measurements of the history of
dust-obscured energy, finding a gradual increase from
$\rm z =0$ to $\rm z=1$. Finally, Viero et al. (2009) investigated the clustering of star
forming galaxies from the power-spectra of the
BLAST maps, and Patanchon et al. (2009) used the distribution of fluctuations in the maps
to estimate the submillimetre number counts.

Whereas these studies looked at the statistical properties of the
BLAST maps or the statistical properties in the BLAST bands
of galaxies in catalogues selected in other wavebands, a sixth paper \citep{dye2009} looked
for the counterparts in other wavebands of the indivdiual sources detected
in the BLAST survey. This is a challenge because the angular resolution of BLAST
(FWHM of 36, 42 and 60 arcsec at 250, 350 and 500 $\mu$m, respectively)
is larger than the angular resolution of SCUBA at
850 $\mu$m (FWHM of 14 arcsec). Nevertheless, using the standard frequentist
technique (\S 2) that has been used for other submillimetre surveys, Dye et al.
(henceforth D09) succeeded in finding radio and/or 24-$\mu$m counterparts
for 227 out of 351 sources detected at 5$\sigma$ 
in the BLAST survey centered on
GOODS-South. 
The authors used the spectroscopic and photometric redshifts that
exist for many of these counterparts to show that 75\% of them
lie at $\rm z < 1$, while only a handful of SCUBA souces lie
at such a low redshift. The luminosities of these counterparts are also lower
than those of the SCUBA galaxies, being more typical of luminous infrared galaxies
(LIRGs) than ULIRGs.
An important point is that the catalogue used by D09 and in this paper is
a 5$\sigma$ catalogue in the sense that $\sigma$ is the instrumental noise, whereas an
additional source of noise is the fluctuations in the map produced by faint sources.
The effect of both types of noise on the fluxes of the sources in the catalogue
is one of the issues we will address in this paper.

This paper represents a continuation of the work described in D09. We present
the results of a redshift survey of the counterparts to the BLAST sources with the
AAOmega multi-object spectrometer on the Anglo-Australian Telescope. We use the
spectroscopic data for a number of different purposes. First, we
use the spectroscopy to investigate the nature of the galaxies that are
the counterparts to the BLAST sources. Second, we use the spectroscopic
data and the imaging data that exists for this field to address the issue of multiple
counterparts to the BLAST sources. This is a familiar problem from attempts
to find counterparts to SCUBA sources \citep{rob2007} and occurs when the frequentist
approach finds multiple possible counterparts to the submillimetre source.
The possible causes are either that the submillimetre
source actually consists of two submillimetre sources confused together---an
obvious strong possibility given the poor angular resolution---or that only one of the
possible counterparts is
a submillimetre source with the second galaxy being physically associated in some
way (possibly
in the same galaxy group) with the first.
Third, we use the spectroscopic redshifts to investigate the accuracy
of the photometric redshifts used in D09. Fourth, we use a combination of spectroscopic
and photometric redshifts to make the 
first estimates of the luminosity function at these wavelengths, and also a first estimate of
the dust-mass function (the space-density of galaxies as a function of dust mass).

The layout of this paper is as follows. In \S 2 we revise the frequentist identification technique
and give a list of secondary counterparts that complements the list of primary counterparts
given in D09. Section 3 describes the redshift survey. Section 4 describes the results of the analysis
based on the redshift survey, including the first estimates of the galaxy luminosity function in this
waveband. Section 5 contains a brief discussion and our conclusions.
We assume everywhere the standard concordance cosmology: $\rm \Omega_M=0.28,\ \Omega_{\Lambda}=0.72,\
H_0 = 72\ km\ s^{-1}\ Mpc^{-1}$.

\section{The Search for Counterparts}

Full details of the multi-wavelength datasets are given in D09.
Briefly, the BLAST survey of GOODS-South consisted of a wide-area map
of 8.7 deg$^2$ with a deeper confusion-limited map of 0.8 deg$^2$. D09
lists a catalogue of all the sources detected at $\rm >5\sigma$ in any
of the three BLAST bands. From the point of view of the detection of
counterparts, there are two distinct regions. The central $\rm 30\times30$ arcmin$^2$
of the deep BLAST survey was surveyed by the Far-Infrared Deep Extragalactic
Legacy Survey (FIDEL) \citep{mag} and is the same region that
was surveyed in the 17-band optical survey COMBO-17 \citep{wolf}.
We call this region the `FIDEL area'. Outside this region, the whole BLAST survey
area has been covered by the Spitzer Wide-area InfraRed Extragalactic Survey (SWIRE)
\citep{lonsdale} in all the Spitzer bands, although only $\simeq$4 deg$^2$
were surveyed by the Spitzer team in optical bands (u,g,r,i,z). The radio catalogues discussed in
D09 also
consist of a deeper central region covering 0.33 deg$^2$ and a wider shallower
catalogue covering $\simeq$4 deg$^2$.

Full details of the identification procedure were given in D09. Here
we revise the main points.
We searched for 24-$\mu$m and radio counterparts to the
BLAST sources using the frequentist approach of Lilly et al. (1999), which
is based on the method of Downes et al. (1986).
The method is to search for possible counterparts close
to the submillimetre position and then use a Monte-Carlo analysis
to estimate the probability that the possible counterpart is there
by chance and is actually not genuinely associated with the
submillimetre source. The advantage in this situation of this frequentist approach over 
Bayesian approaches \cite{will} is that it does not require much
information about the positional errors, which in this case
are poorly known because of the effects of source confusion.
The details of the procedure are given here for a radio
catalogue but are the same for a 24-$\mu$m catalogue.

\begin{enumerate}

\item Select a random position within the area common to the BLAST and radio
catalogues.

\item Find the minimum of the quantity $S = r_{sep}^2 n(>f)$, where $r_{sep}$ is
the separation between a radio source and the random position, $f$
is the flux density of the radio source, and $n(>f)$ is the
surface density of radio sources brighter than this radio source. An important
point is that only radio sources within a maximum separation radius of $r_{max}$ are included
(see below).

\item Repeat steps 1 and 2 for N realizations to determine the distribution of $S$ for
the radio sources.

\end{enumerate}

We determined this distribution separately for the FIDEL area and the area
outside FIDEL.
If we then have a real potential counterpart with a value for $S$ of $S_i$, we can 
estimate the probability that it is simply there by chance from the distribution of
$S$ generated by the Monte-Carlo simulation, $D(S)$. The probability that the potential counterpart
is simply there by chance is

\smallskip
$$
P(S<S_i) = {1 \over N} \int^{S_i}_0 D(S) dS. \eqno(1)
$$
\smallskip

The crucial point that was investigated in D09 was the choice of $\rm r_{max}$. Even though
the accuracy of the submillimetre positions is uncertain, we do know enough about the accuracy of the
positions to insist on a maximum value for $r_{sep}$; otherwise 
a bright radio source such as Cygnus A would
yield a low value of $P$ even if it were many degrees away from the BLAST source. The choice of the
value of $r_{max}$ is a balance between not missing genuine counterparts and including too many false
IDs. D09 describes a method  for determining the value of $r_{max}$ at which the expected number
of excluded genuine counterparts equals the number of included false counterparts. A byproduct
of this analysis was an estimate of the distribution of offsets between the positions
of the BLAST sources and the counterparts.

\smallskip
$$
n(r) \propto r e^{-r^2 \over 2 \sigma^2} \eqno(2)
$$
\smallskip
\noindent with $\sigma \simeq$8 arcsec. 
This agrees well with a prediction (D09) based on 
the analytical formula for the positional errors of submillimetre
sources derived by Ivison et al. (2007).
We derived a value for $r_{max}$ of 20 arcsec in the FIDEL area
and a value of 25 arcsec outside this area.

In D09 we listed the counterparts with $P < 0.05$ for the BLAST sources in the FIDEL area and the counterparts
with $P < 0.1$ for the BLAST sources outside this area. The different values of $P_{max}$ in the two regions were
chosen because of the different surface densities of 24-$\mu$m and radio sources in the two
regions. By summing the values of $P$ for our list
of 227 posible counterparts, we predict that $\simeq$5 are incorrect.
The counterparts
listed in D09 were the primary counterparts, the counterpart to each BLAST source that had the lowest value
of $P$ and satisfied the condition $P < P_{max}$. In Table 1 we list  
the counterpart with the next lowest value of $P$ and
$P < P_{max}$, if one exists.
There are 69 of these `secondary counterparts' compared to 227 primary counterparts. Approximately
one third of the BLAST sources have more than one possible counterpart. 
Fig. 1 shows the optical or mid-infrared images of all
BLAST sources with more than one counterpart. We will present images of all the counterparts in a
later paper (Moncelsi et al. in preparation).
For the sources in the FIDEL region these images are taken from the optical R-band image from the COMBO-17
survey (Wolf et al. 2004). For the sources outside the FIDEL region, the image is taken by preference
from  an image taken in the r-band by the SWIRE team (Lonsdale et al. 2004) and, if that does not exist, from the
Spitzer 3.5-$\mu$m image. We discuss the reason for multiple counterparts in \S 4.

\section{The Redshift Survey}

On 24th November 2008 we used the AAOmega spectrometer on the Anglo-Australian Telescope to observe
targets from a preliminary list of counterparts to the BLAST sources. AAOmega \citep{sharp}
consists of 392 fibres
that feed the light from targets within a field 2 degrees in diameter to
a blue and a red camera via a dichroic. We used the 580V and the 365R gratings for the blue and the red cameras
respectively, which 
gave a wavelength coverage from 370 to 880 nm
and a resolution ($\lambda/\delta \lambda$) of 1300.

We adopted the following scheme for placing the fibres. We only placed fibres on targets
with sufficiently accurate positions (the fibres are 2.1 arcsec in diameter).
If a BLAST source had a radio counterpart, we 
placed the fibre on the radio position. If 
a BLAST source had only a 24-$\mu$m counterpart, we searched for an optical or
3.6-$\mu$m counterpart within 3 arcsec of the 24-$\mu$m position; we
placed the fibre on the optical position if an optical counterpart existed and, if not,
on the
the 3.6-$\mu$m position. We were constrained in our placement of fibres
by the geometry of the BLAST survey area. The wide-area survey is much larger
than the field-of-view of AAOmega, whereas the deep central area in which there is the
greatest density of counterparts is significantly smaller (0.87 deg$^2$) than the AAOmega field
of view. To observe our main target list, we used three configurations of fibres.
In each configuration, we observed the same targets in the BLAST deep area but a
different set of targets in the surrounding area, observing the central targets with
12 exposures of 1800 seconds and the surrounding targets with 4 exposures of 1800 seconds.
There were far more fibres than counterparts to 5$\sigma$ BLAST sources with accurate positions,
and so this list consisted of counterparts to BLAST sources with signal-to-noise $\rm >3.5$.
Even then we had spare fibres, so we placed the remaining fibres on the counterparts to
sources detected in the SWIRE survey (Lonsdale et al. 2004). The counterparts to the BLAST sources
in this list were always the primary counterparts, the ones with the lowest probability
of being there by chance. We also observed a second  list of targets, which contained
many of the secondary counterparts (\S 2) but we only succeeded in observing this list for two exposures
of 1200 seconds (we had planned to come back to this field on succeeding nights but the rest of the
run was lost because of the weather).

We reduced the data using the standard data reduction pipeline for AAOmega, 2dfdr,
which is described on the
AAOmega website\footnote{http://www.aao.gov.au/AAO/2df/aaomega/aaomega.html}. We then extracted redshifts from the spectra using the RUNZ software package
(Croom, private communication). This 
package automatically extracts a redshift from each spectrum by fitting continuum templates
to the spectrum and by looking for emission lines. A crucial aspect of the programme is that the
user is able to inspect the result and use his/her judgement to, if necessary, change the redshift, for
example if one of the lines used in the fit looks like an artefect.
Two of us (SAE and LM) did this independently and agreed in all cases.
We obtained spectroscopic redshifts for 399 galaxies, 82 of which are
primary or secondary counterparts to 5$\sigma$ BLAST sources.
Out of the 669 targets in our main list, a mixture of BLAST and SWIRE galaxies,
we obtained spectroscopic redshifts
for 339,
a success rate of 51\%.

Table 2 contains the list of spectroscopic redshifts we have obtained
for the counterparts to the 5$\sigma$ BLAST sources. 
The counterparts are mostly the
primary counterparts but nine are the secondary counterparts. The other redshifts, which are
for
a mixture of SWIRE galaxies and counterparts to BLAST sources
that now fall below the 5$\sigma$ cutoff, are given in Table 3. 
Figure 2
includes spectra of a representative sample of the BLAST counterparts.

It is impossible to quantify accurately the probability of a redshift
being correct, because the final redshift is a combination of
the automatic continuum and emission-line fitting by RUNZ plus the
subjective judgement of the user. We adopted the
quality assessment system used by RUNZ, in which a redshift
with a quality of 5 is defined as being a `definite redshift', one with
a quality of 4 is defined as being `almost certain with roughly a 95\% probability
of being correct' and one with a quality of 3 as being `somewhat less certain
but probably correct'. We have listed in Tables 2 and 3 our estimates of the
quality of each
redshift using this system. In many cases, the value is simply the one
produced by RUNZ; in others it is our modification of the RUNZ value based
on an inspection of the spectrum.
Of the 82 redshifts we measured
for the BLAST counterparts, two have a quality flag of 3, ten have a quality flag
of 4 and the remainder have a quality flag of 5.

Taylor et al. (2009) have measured spectroscopic redshifts for 21 of the same
galaxies for which we have measured redshifts. In 19 cases, our redshifts
and those of Taylor et al. agree. In the case of the first discrepancy,
our redshift is 0.672, whereas that of Taylor et al. is 0.553. We inspected our
spectrum and the redshift has a quality flag of 5 and
is based on detection of an [OII] 372.7 emission line and
several absorption features. We are therefore confident our redshift is
correct. In the case of the second discrepancy, our redshift is a poor quality
one (quality flag of 3) of 0.205, whereas the redshift of Taylor et
al. is 0.620. Our redshift is mostly based on a fit to the
continuum and a single emission line, which we have assumed
is [OIII] 500.7. Taylor et al. have clearly assumed the
line is [OII] 372.7. On reinspection of our spectrum, we decided
this is more likely to be correct, and so the redshift we have listed is based
on this assumption.

\begin{figure*}\figurenum{3}
\plotone{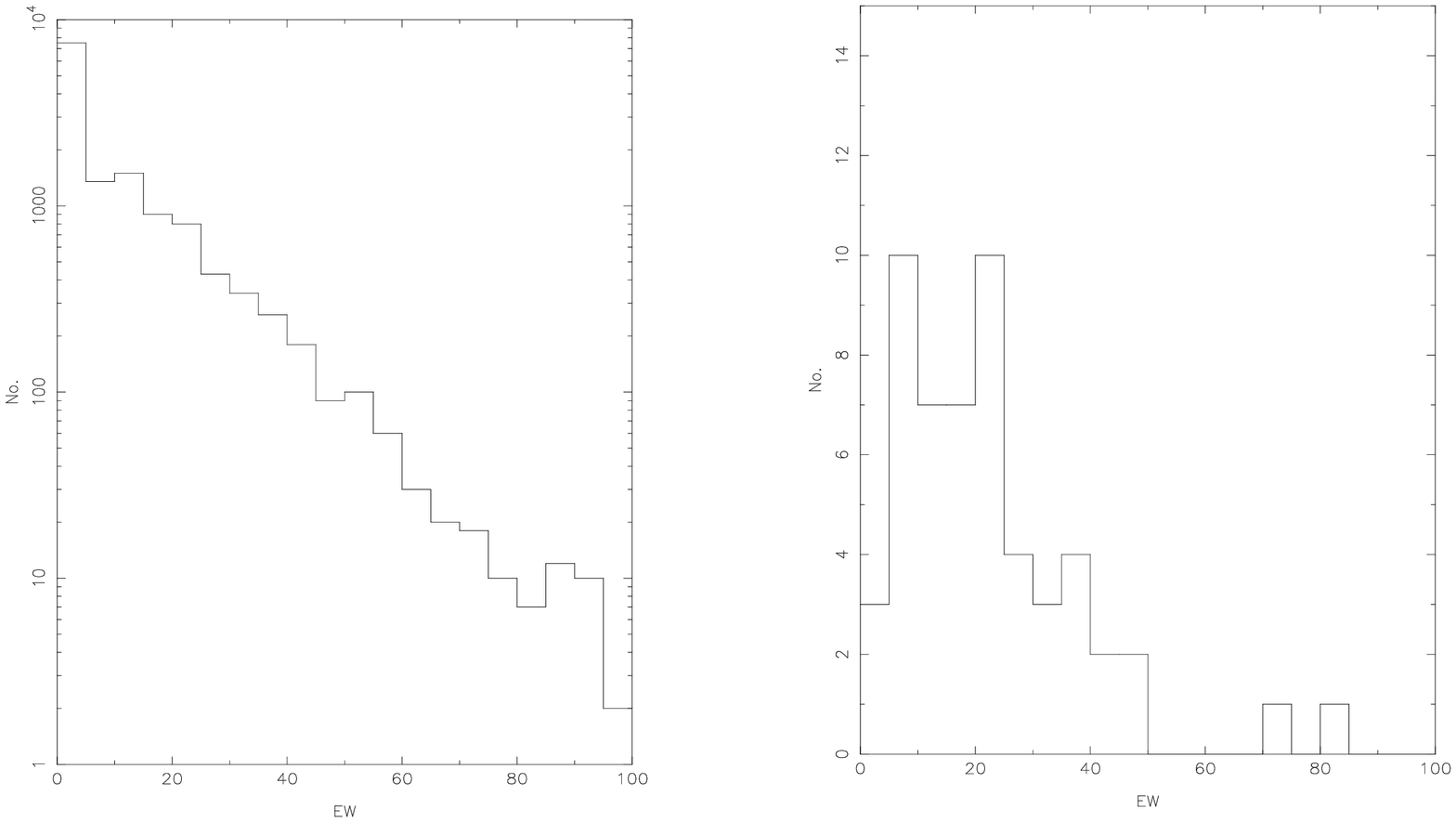}\caption{Histograms of the equivalent widths of the H$\alpha$ line for
a sample of 25,000 galaxies selected from the Sloan
Digital Sky Survey and the 2dF Galaxy Redshift Survey by
Balogh et al. (2004; left panel) and for
the BLAST primary counterparts (right panel). Note that
the y-axis in the left panel is on a logarithmic scale but
not the one in the right panel.
}
\end{figure*}

\section{Results}

The spectra in Fig. 2 are a representative sample of the spectra we obtained for the
BLAST counterparts, consisting of spectra of the 1st, 6th, 11th etc. galaxies
listed in Table 2. They show quite clearly that 
we were undoubtedly more
successful in obtaining spectroscopic redshifts for galaxies with low redshifts than for galaxies
at high redshifts. It was fairly easy to measure the redshift of a galaxy
at
$\rm z < 0.3$ because of the presence of the bright H$\alpha$ and [NII] 658.3 lines, 
the [SII] 671.6, 673.1 doublet and often many other absorption and emission lines.
At $\rm 0.3 < z < 1.0$, it was a little more difficult: the H$\alpha$ and
[NII] 658.3 lines are redshifted out of the accessible waveband, and the only bright emission line is
the [OII] 372.7 line. In this range of redshifts, the redshift usually came from this line plus a fit to
the continuum. At even higher redshifts, the [OII] line is redshifted out of the waveband, and
it was only possible to extract a redshift if the object is a quasar because quasars have several
broad emission lines, such as $\rm CIII] 190.9$, which appear in the accessible waveband
at $\rm z > 1$. Thus the success rate for
obtaining redshifts for BLAST counterparts that do not have active nuclei probably falls from close to
100\% at $\rm z=0$ to close to 0\% at $\rm z \simeq 1$. Therefore, for investigating the evolution
of the luminosity function (\S 4.4), we are still reliant on photometric redshifts.

\subsection{Inferences from the Spectra}

The absolute flux scale of our spectra and its dependence on wavelength
is uncertain, since we did not observe spectrophotometric standards.
Nevertheless, we were able to measure two quantities from the
spectra that do not require this:
the equivalent width of the H$\alpha$ line
and the ratio of the flux in the [NII] 658.3 line to the flux in the
H$\alpha$ line, two lines which are very close in wavelength.
In Table 2 we have listed the H$\alpha$ equivalent width of each
galaxy, corrected to the galaxy's rest frame, and the value of this
line ratio. If there are no values for these
quantities, it is either because the galaxy is at too high a redshift 
for these to be measured or
because there was a problem with the spectrum
or because the H$\alpha$ line is broad.

The H$\alpha$ equivalent width is useful because it gives a measure of the
star-formation rate in the galaxy relative to its average rate since the
galaxy was formed. Figure 3 shows a histogram of the H$\alpha$ equivalent width
for the BLAST primary counterparts and for
a sample of $\simeq$25000 galaxies with $\rm 0.05 < z < 0.095$ drawn from
the Sloan Digital Sky Survey (SDSS)  and the 2dF Galaxy Redshift Survey (2dFGRS; Balogh et al.
2004).
Note 
that because of the very large number of galaxies with equivalent widths less
than 5 $\AA$ in the latter sample, the plot for this sample is on
a logarithmic scale.
These galaxies with low equivalent widths represent the `old, red and dead' population
in which the star-formation rate was much higher in the past than it is today.
It is now clear that there is a dichotomy between the properties
of this population and those of actively star-forming galaxies \cite{kauffmann}.
The most striking difference between the
two samples is that there are very few BLAST galaxies with equivalent widths less
than 5 $\AA$, and so the BLAST counterparts are almost exclusively drawn from
the actively star-forming population.  
However, if we exclude the galaxies in both samples with equivalent widths $\rm < 5 \AA$,
the two samples appear quite similar, with the mean equivalent width being 20 $\AA$ for the
SDSS/2dFGRS sample and 24 $\AA$ for the BLAST sample. Therefore, judged
by the H$\alpha$ equivalent width, the BLAST counterparts are star-forming galaxies
but do not appear to be extreme ones. We will address the issue of whether the BLAST galaxies
are exceptional or run-of-the-mill galaxies in more detail in a subsequent paper (Moncelsi et
al. in preparation).

Five of the BLAST counterparts clearly contain powerful active galactic nuclei (AGN) because they have
spectra typical of quasars. We looked for less powerful active nuclei by measuring line ratios.
The classic way of determining whether the emission lines from a galaxy are dominated by emission from gas
that is photoionised by an active nucleus or by gas that is heated by young stars is to look at the galaxy's
position on a line ratio diagram. For example, the two classes 
fall in separate regions in a diagram of [NII] 658.3/H$\alpha$ verses [OIII] 500.7/H$\beta$ \citep{bpt81}. We almost
never had measurements of all four of these lines, but we did have measurements of the first ratio for
many of the galaxies. In their study of the galaxies detected in the Sloan Digital Sky Survey (SDSS), Miller et
al. (2003) argued that
if this line ratio is $>$0.63 the galaxy must lie in a region of the four-line diagram dominated by AGN (the
reverse is not true because if the line ratio is below this value, the galaxy may still lie
in a region of the diagram that is dominated by AGN). We were able to measure this ratio for 54 of
our sample of 73 primary counterparts for which we have redshifts. 
If we include the objects with a value for this line ratio $>$0.63, the quasars and two other
objects for which there is some evidence for an AGN (broad H$\alpha$ in one case and a large value
for the ratio of the [OIII] 500.7 and H$\beta$ lines in the other), there is evidence for
an AGN in 19 out of 73 primary counterparts. This percentage (26\%) is rather greater
than the percentage of SDSS galaxies that contain AGN, which is approximately 18\% 
\citep{miller}, but given the differences between how these samples were selected we do not
believe we can draw any profound conclusions from this result.

\begin{figure*}
\figurenum{4}
\plotone{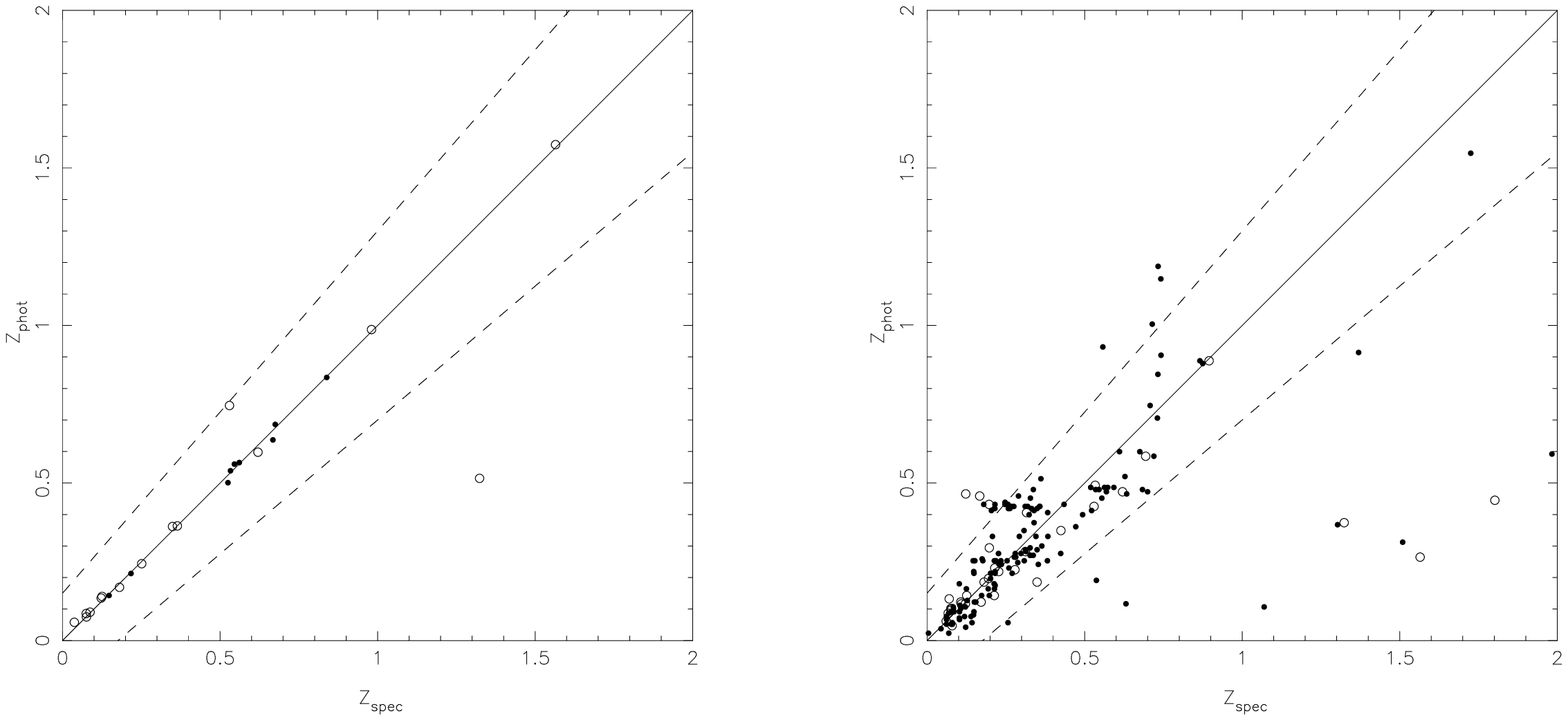}
\caption{Plot of spectroscopic redshift verses redshift
estimated using a photometric redshift method. In the left-hand
panel, the photometric redshifts were estimated as part of the
COMBO-17 survey \citep{wolf} and in the right-hand panel
they were estimated by Rowan-Robinson et al. (2008) from
broad-band optical and Spitzer photometry. In both panels,
the open circles are galaxies that are primary counterparts
to BLAST 5$\sigma$ sources; the crosses are other galaxies that
were detected in the Spitzer SWIRE legacy survey \citep{lonsdale},
many of which were also detected by BLAST.
The continuous line shows where the photometric and spectroscopic
redshifts are equal. The dashed lines show the limits, beyond which
the photometric redshifts are classified as `catastrophic errors'.
Once objects flagged as quasars are removed from the right-hand panel (see text),
the percentage of catastrophic errors is $\simeq$9\% for both methods. After
removing the catastrophic errors, the redshift errors (see text
for definition) are 0.031 for COMBO-17 and 0.056 for
SWIRE.
}
\end{figure*}

\subsection{A Test of Photometric Redshifts}

We have not measured enough spectroscopic redshifts to be able to investigate cosmic evolution without recourse
to photometric redshifts. A potential problem with using photometric redshifts is that the
BLAST galaxies probably contain large amounts of dust, and most photometric redshift techniques
have only been tested on galaxies detected in optical surveys. The only method that has been
tested on galaxies that may be similar to those detected by BLAST is that
of Rowan-Robinson et al. (2008), who estimated redshifts for
galaxies detected in the Spitzer SWIRE Legacy
Survey. 

We used the spectroscopic redshifts to test the accuracy of the two sets
of photometric redshifts that we used in our investigation of the
evolution of the luminosity function (\S 4.4). The first set were obtained
by Wolf et al. (2004) from the COMBO-17 survey, a survey of a field $\rm 30 \times 30$ arcmin$^2$ in size through
17
optical filters. The second set
were obtained by
Rowan-Robinson et al. (2008) 
for the much larger area ($\simeq$4 deg$^2$)
covered
by the optical images taken as part of the SWIRE survey. 
The quality of the imaging data used to estimate the
second set of redshifts (three broad-band optical images plus the Spitzer
images at 3.6 and 4.5 $\mu$m) is more typical of the imaging data
available over large areas of sky. Since there are many
new large-area
surveys, such as
those with Herschel
and VISTA \citep{will2}, which will be reliant on
photometric redshifts, a test of the the latter set
is particularly interesting.
Our test is unbiased because our new spectroscopic redshifts are not part of the
sets that were used to tune either of the original methods.

Figure 4 shows a comparison of the photometric and spectroscopic redshifts for the two datasets. We have included
all the galaxies for which we have spectroscopic redshifts (i.e. Tables 2 and 3).
We have only included
spectroscopic redshifts with a quality of four or greater (\S 3), to ensure that any discrepancies are
caused by errors in the photometric redshifts rather than errors in the spectroscopic ones. 
We only included SWIRE photometric redshifts that were based on photometry in at least
three optical bands (Rowan-Robinson et al. 2008).
In our error analysis, we used $\delta = { z_{phot} - z_{spec} \over 1 + z_{spec} }$
as our measure of the discrepancy between the photometric and spectroscopic redshift,
and we treated any photometric redshift with $\rm |\delta| > 0.15$ as a catastrophic error.

We found that the percentage of catastrophic errors is $\simeq$8\% for the COMBO-17 photometric
redshifts and $\simeq$15\% for the SWIRE redshifts. 
However, many of the discrepant SWIRE photometric redshifts
are for objects flagged by Rowan-Robinson et al. (2008) as probable quasars (confirmed
in 9 out of 10 cases by our spectroscopy and probably
also true in the tenth case), and once these objects are
removed the percentage of catastrophic errors falls to 9\%.
The very discrepant object in Figure 4a
was also shown by our spectroscopy to be a quasar but was not flagged as
such in the COMBO-17 survey.
Excluding all objects for which there
are catastrophic errors, we estimate that the errors for the
two methods
($\sqrt{<\delta^2>}$) are 0.031 for the COMBO-17 dataset and 0.056 for the the
SWIRE dataset. In calculating the luminosity function (\S 4.4),
we have used redshift bins with a width of 0.2 in redshift, and 
we therefore conclude that the errors in the photometric redshifts are not
likely to be the limiting factor in the accuracy of our analysis.

\subsection{Multiple Counterparts}

A common problem with submillimetre surveys is that there are sometimes two possible counterparts
to the submillimetre source, each of which has a low probability of being there
by chance. Ivison et al. (2007) found that approximately 10\% of sources detected
in the SHADES 850-$\mu$m survey have more than one counterpart. For the BLAST
survey of GOODS-South, the percentage of sources with multiple counterparts
is $\simeq$30\% (\S 2). Of the 69 sources with multiple counterparts,
13 have more than two counterparts, although, as we show below, in three
cases the 24-$\mu$m/radio sources are all associated with a single large
galaxy.  

\begin{figure}
\figurenum{5}
\plotone{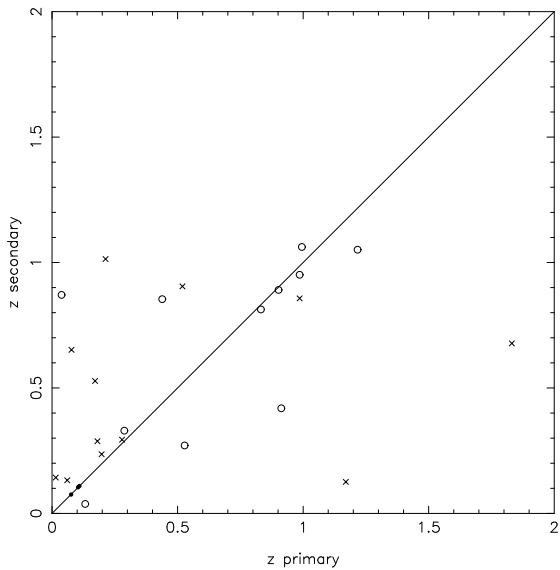}\caption{Plot of redshift of the primary countpart 
verses theredshift of the secondary counterpart for the 
BLAST sources for which there are two counterparts. The filled
circles represent sources for which there are spectroscopic
redshifts for both counterparts; the crosses represent sources
for which there is one spectroscopic redshift and one photometric
redshift; and the open circles shows sources for which both counterparts
have photometric redshifts. The straight line shows where the redshifts
of the two counterparts are the same.
}\end{figure}

This high percentage might, of course, be due to
the clustering of 24-$\mu$m or radio sources in the catalogues used
for the identification analysis; one of the counterparts might
be
genuine with the second simply being there
because of the clustering within the radio/24-$\mu$m catalogue.
We have investigated this by adapting
the Monte-Carlo simulation described in D09. We lay down
points randomly in the area covered by the 24-$\mu$m and radio
catalogues,
carrying out separate simulations for the radio and 24-$\mu$m catalogues
and for the area covered by the FIDEL
survey and for the area outside this survey (\S 2). The points represent artificial
submillimetre sources. We then apply the frequentist identification
technique to look for counterparts at this list
of positions, using the same search radii that
we used for the real data. For all the counterparts that satisfy
the condition that the probability is $\rm <P_{max}$, we determine the
percentage of cases in which there is also a second counterpart that
satisfies this condition. We find that the percentages in the FIDEL region
are 10.7\% for the radio catalogue and 3.3\% for the
24-$\mu$m catalogue, with the percentages outside the FIDEL
area being 7.2\% for the radio catalogue and
5.8\% for the 24-$\mu$m catalogue.
These percentages should also be good estimates of the percentages
of the real counterparts that have secondary counterparts 
because of clustering in the radio/24-$\mu$m catalogues.
Applying these percentages to the real list of counterparts,
we estimate that $\simeq$15 of the 69 sources with multiple counterparts
are caused by this effect. Therefore the true number of sources
with multiple counterparts is $\simeq$54, $\simeq$24\% of the total
number of sources with counterparts. This is still larger
than the percentage found by Ivison et al. (2007) for
the SHADES survey, which is
not surprising
because of the poorer angular resolution of BLAST   
(36 arcsec at 250 $\mu$m verses 14 arcsec
for SHADES), although the difference in effective linear resolution is
somewhat less because the BLAST sources tend to be at lower redshift (D09);
for example, 
the linear resolution for a BLAST source at $\rm z = 0.3$ is only 33\% greater
than the
linear resolution for a SHADES source at $\rm z = 2$. 

There are a number of possible explanations of multiple
counterparts. One of these
is the possibility that there is a single genuine counterpart, which
has been gravitationally lensed by a nearby galaxy \citep{blain2}.
The arguments against this are (i) the angular distances between
the counterparts are often much greater than the typical distances
between lensed source and a lense and (ii) that the BLAST galaxies
are typically spirals or interacting galaxies, whereas most lenses
are predicted to be ellipticals \citep{blain2}.
There are two explanations that cannot be ruled out:

\begin{enumerate}

\item The two counterparts are physically associated in some
way. They might either be in the same cluster or two galaxies
that are gravitationally interacting. If this explanation is
correct, the counterparts need not both be submillimetre sources. 

\item Both of the counterparts are submillimetre sources which 
are not linked
physically in any way.

\end{enumerate}

We can investigate which of these is correct
by, first,
inspecting the images of all the BLAST sources with multiple counterparts and,
second, by comparing the redshifts of the primary and secondary counterparts.
The images (Fig. 1) immediately reveal a few interesting results. First, there
are several sources where there is clearly only a single galaxy. In these
cases, it seems almost certain that the two apparent counterparts are actually
radio or 24-$\mu$m sources within the galaxy. For example, BLAST 4 (see Table
1 for full name) has two radio
counterparts which lie close to the centre of a large galaxy. Inspection of the
radio image shows that these two sources are actually two peaks in a single
source that is extended in the same direction as the optical structure. Not surprisingly,
our spectroscopic redshifts of these two counterparts are virtually the same. BLAST 2
and BLAST 53 also seem to be cases where the two apparent counterparts
are actually sources within a single galaxy. Second, there are several cases
where the counterparts seem to be clearly interacting. In four cases---BLAST 6,
9, 103 and 127---there is clear morphological evidence for
a gravitational interaction between the two counterparts.

Unfortunately, this still leaves a large number of multiple counterparts
for which there are no clear morphological clues. We can make
more progress by comparing the redshifts, either spectroscopic
or photometric, of the primary and secondary counterparts.
We have redshifts for
both counterparts for 27 systems, although unfortunately spectroscopic
redshifts for both counterparts in only three cases. Fig 5 shows 
there is
a clear correlation
between the two redshifts (Spearman's $\rho$ = 0.52; probability of the two
variables being uncorrelation is $\simeq$0.4\%) with 14 out of the 27 systems lying sufficiently
close (given the errors on the photometric redshifts)
to the line on which the redshifts of the two counterparts are the same. 
Therefore, since
there is no reason why there should be a correlation if the
second explanation is correct,
we conclude that in at least half the cases where there are multiple counterparts,
the counterparts appear to be physically associated. 

\subsection{Luminosity Functions}

In this section we make a first attempt to estimate luminosity functions at the three BLAST wavelengths.
There are two major obstacles to overcome. First, there are still many BLAST 5$\sigma$ sources
that do not have counterparts, and some of the counterparts do not have redshifts. Second,
the fluxes of the BLAST sources are systematically biased upwards by the effect of noise,
both instrumental noise and the fluctuations in the map caused by other submillimetre sources.
This is the well-known Eddington bias
\citep{edd}, in which the effect of the deep differential source counts
is that more sources in a flux-limited sample have had their fluxes increased by
noise than decreased by noise. 
The effects of Eddington bias in the BLAST maps is why our earlier papers
concentrated on the statistical properties of the BLAST maps rather than
the properties of individual sources (e.g. Patanchon et al. 2009). Here, if
we wish to use the information about cosmic evolution provided by our redshift survey,
we are forced to confront its effects.
The confusion of sources discussed in \S 4.3 is often
treated as a separate problem, but this is really  just a form
of Eddington bias in which the noise comes from discrete sources. This upwards
bias in the fluxes of sources is often called `flux-boosting'.
We have developed a number of techniques for overcoming
these obstacles and believe that our conclusions at the end of
this section are not 
invalidated by any of these effects.

For each wavelength, the sample from which we start is the sample of sources detected at 5$\sigma$ at this
wavelength and which fall in a 4.2 deg$^2$ area covered by either the SWIRE optical images or the COMBO-17 survey.
The point of restricting the investigation in this way is that outside
this area there are no photometric redshifts. The three samples are thus subsamples of the list
in Table 3 of D09. Table 4 in this paper lists the statistics of these samples: the number of sources; the 
number with either a radio or a 24-$\mu$m counterpart or both;
the number with either a photometric redshift or a spectroscopic redshift, with the number of spectroscopic
redshifts in brackets. The table illustrates one of the problems mentioned above, that not
all the sources have counterparts and some of the counterparts do not have redshifts, either spectroscopic
or photometric. 

We will address the problem of the lack of redshifts first because
it is easiest to deal with. It seems likely
that counterparts without redshifts are galaxies that are at very high redshift, and are thus
too faint at optical wavelengths for photometric redshift methods to work. We can test this
by comparing the mid-infrared colours of the counterparts with and without photometric
redshifts. Figure 6 shows all the counterparts on a plot of
$\rm S_{3.6}/S_{4.5}$ verses $\rm S_{5.8}/S_{8.0}$, colors which
Pascale et al. (2009) show depend on redshift. The counterparts
without photometric redshifts cluster in the bottom right of the figure, which
Figure 3 of Pascale et al. shows corresponds to $\rm z > 1$. Therefore,
we can assume that as long as we restrict our estimates of the luminosity function
to $\rm z < 1$, our estimates should not be affected if we omit these objects.

\begin{figure}\figurenum{6}
\epsscale{.60}\plotone{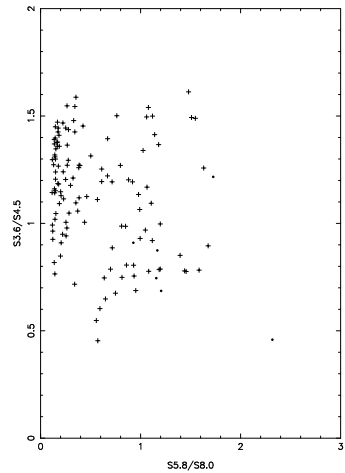}\caption{Plots of the 
ratio of 3.6 $\mu$m to 4.5 $\mu$m flux density
verses the ratio of 5.8 $\mu$m to 8.0 $\mu$m flux density for all
the primary counterparts for which there are measurements of these 
flux densities. The crosses are counterparts for which there are redshifts, 
either spectroscopic or photometric; the filled circles are counterparts for
which there is no redshift measurement or estimate.}
\end{figure}

The problem of the missing counterparts is more complicated because there are several possible causes. The first
is that the BLAST sources without counterparts are at such high redshifts
that their 24-$\mu$m and radio fluxes fall below the limits of the 24-$\mu$m and radio catalogues.
Evidence for this is the fact that the percentage of BLAST sources with counterparts falls with
increasing wavelength, in line with the predictions of models that the fraction of sources
at very high redshift should increase with increasing wavelength (D09). If this is the cause,
then we should again have no problems if we restrict our estimates of the luminosity function to low redshifts.
The second possibility, suggested by the simulations in Appendix 2, is that some of these
sources are not genuine sources
or are a confused combination of instrumental noise and many very dim sources.
If these sources are not real sources, of course, omitting them from
our estimates of the luminosity function is the right thing to do.
However, a third possible explanation, which
must be correct for at least some of the BLAST sources, is that the
24-$\mu$m or radio counterpart is in the catalogue but our frequentist method of finding
the counterparts (\S 2) has failed to find it. It is possible to estimate
the number of counterparts that are present in the catalogues but are missed by the
selection procedure from the  
predicted distribution of distances between the BLAST sources
and their
counterparts that was given in equation 2.
Each counterpart found by the
method should be multiplied by a correction factor to compensate for the counterparts that
could not have been found by our frequentist method. This is
given by

\smallskip
$$
c_i = { \int^{\infty}_0 r e^{-r^2 \over 2 \sigma^2} dr \over \int^{r_{cut,i}}_0 r e^{-r^2 \over 2 \sigma^2} dr} \eqno(3)
$$
\smallskip
\noindent in which $\sigma\simeq$ 8 arcsec (\S 2) and
$r_{cut,i}$ is the maximum radius at which this counterpart could have been found
by this method. For counterparts that are bright 24-$\mu$m or radio sources,
$r_{cut,i}$ is the same as the search radius, $r_{max}$. 
However, a faint 24-$\mu$m source would have been dismissed as
a possible counterpart at a smaller distance from the BLAST source
than $r_{max}$
because its probability of being there by chance would have exceeded $P_{max}$.
Therefore, 
$r_{cut},i$ is given by the smaller of the search radius, $r_{max}$, and
\smallskip
$$
r' = ({P_{max} \over P_i})^{1/2} r_i \eqno(4)
$$
\smallskip
\noindent in which 
$r_i$ is the distance of the counterpart from the
BLAST position and $P_i$ is the probability that the potential counterpart is not genuinely associated with the
BLAST source.
The number of counterparts missed by the selection
procedure is given by:
\smallskip
$$
N_{miss} = \sum_{i=1}^n c_i - n \eqno(5)
$$
\smallskip
\noindent in which $n$ is the number of counterparts. 

\begin{figure}
\figurenum{7}
\epsscale{1.0}
\plotone{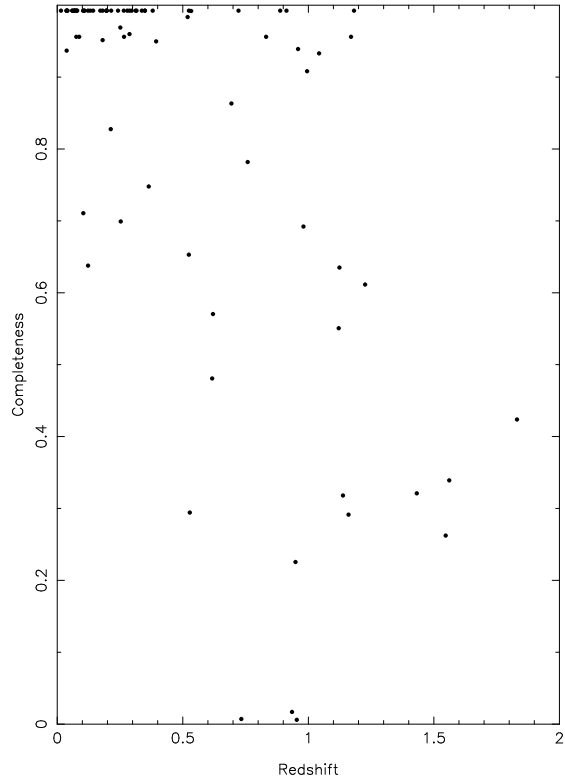}\caption{Plot of the completeness 
(the inverse of the correction factor, \S 4.4) 
verses redshift for all primary counterparts for which there is either
a photometric or a spectroscopic redshift.}\end{figure}

Fig. 7 shows a plot of $1/c_i$ verses redshift
for all counterparts of 250-$\mu$m sources with
either photometric or spectroscopic redshifts. There is a small correction factor
for all counterparts simply because we have used a maximum search radius ($r_{max}$) in the procedure,
and so some true counterparts will have been missed no matter how bright
they are at radio wavelengths or at 24 $\mu$m. However, for some
counterparts the correction factor is much larger because they are faint 24-$\mu$m/radio sources and therefore
only have sufficiently low values of $P_i$ if they are very close to the BLAST position. The fraction of counterparts
with high values of $c_i$ increases with redshift, showing this incompleteness effect is worse
at higher redshifts. The number of missing counterparts, according to
equation 5 is 388, which is clearly too high because the number
of missing counterparts in Table 4 is only 21. However, the number of missing counterparts
is dominated by
the three counterparts with extremely high values of $c_i$. If we omit these as statistical fluctuations,
we obtain a value of 24. Therefore, although the number of counterparts missed by the method
is clearly uncertain, partly because of the effect of huge statistical fluctuations
caused by small offsets and partly because the value of $\sigma$ we have
used in equation 3 has a large error,
it is clear that some of the missing counterparts to the
250 $\mu$m sources must be
in the radio/24-$\mu$m catalogues. We will return to how we correct for
these missing counterparts after discussing the best way of estimating the
luminosity function.

The standard method of estimating the luminosity function is fairly simple. Suppose one
wishes to estimate the value of the luminosity function (the space-density of galaxies
as a function of luminosity) in a particular range of redshift and luminosity. If there
are $n$ galaxies in this luminosity-redshift bin, the standard estimate of the value of the luminosity
function in this bin is:
\smallskip
$$
\phi(L_1<L<L_2,z_1<z<z_2) \Delta log_{10} L = \sum_{i=1}^n {1 \over V_i} \eqno(6)
$$
\smallskip
\noindent in which $V_i$ is the comoving volume in which the i'th galaxy could both have been
detected by the survey and still have been found within the range of redshifts
for this bin. The error on this estimate is usually given as ${1 \over\sqrt{n}}$,
athough this is only an approximation because the values of $V_i$ for the 
sources are usually different.

However, we have preferred to use a different estimate of the luminosity
function which has several major advantages for deriving the luminosity
function from submillimetre surveys. In this method, which was
first
suggested by Page and Carrera (2000), the luminosity function is given by:
\smallskip
$$
\phi(L_1<L<L_2,z_1<z<z_2) \Delta log_{10} L = {n \over V} \eqno(7)
$$
\smallskip
\noindent in which $n$ is again the number of galaxies in the
bin and $V$ is the accessible
comoving volume averaged over
the luminosity range in this bin. 
The important difference is that $V$ is now not calculated from
the measured luminosities of the sources. Page and Carrera have shown
that this estimator is always better than the one given in equation (6), 
and the error on the estimate is now truely ${1 \over\sqrt{n}}$.
A major advantage for submillimetre surveys is that whereas $V_i$ in equation
6 depends on the luminosities of the sources, which are often
uncertain because of flux-boosting, $V$ in equation 7 does not depend
on the measured luminosities of the sources.
This is not a complete solution for the problem of flux-boosting,
which we will discuss more below, because the number of sources, $n$,
in a bin obviously depends on their luminosities having been measured
correctly; but both the
methods suffer from this problem, while only the standard method
suffers from the problem that an error in the luminosity of a source
also produces an uncertainty in $V_i$.
$V$ in equation 7 is given by
\bigskip
$$
V = {1 \over \Delta log_{10}(L)} \int^{L_u}_{L_d} \int_{survey} \int^{min\{z_u,z(L,S_{min}(A))\}}_{z_d}$$ \newline
$${c \over H_0} {D^2 \over \sqrt{\Omega_M (1+z)^3 + \Omega_{\Lambda}}} dz dA dlog_{10}(L) \eqno(8)
$$
\bigskip
\noindent in which $\Delta log_{10}L$ is the width of the bin in luminosity; $dA$ is an element of the BLAST survey area;
$\rm S_{min}$ is the minimum flux density a galaxy could have and still be detected in this area 
element\footnote{A subtle and important point is that in calculating $V$ we do not have to take
any account of flux boosting. In calculating $V$, wherever our model
galaxy is in the survey region, 
the probability of its flux being increased by the effect of noise, whether instrumental or
from the fluctuations of faint sources, is the same as the probability of its
flux being decreased by the effect of noise.};
$L_d$, $L_u$, $z_d$ and $z_u$ are the limits of the bin in luminosity and redshift; and all the cosmological
terms
have their usual meanings. 

\begin{figure*}\figurenum{8a}\plotone{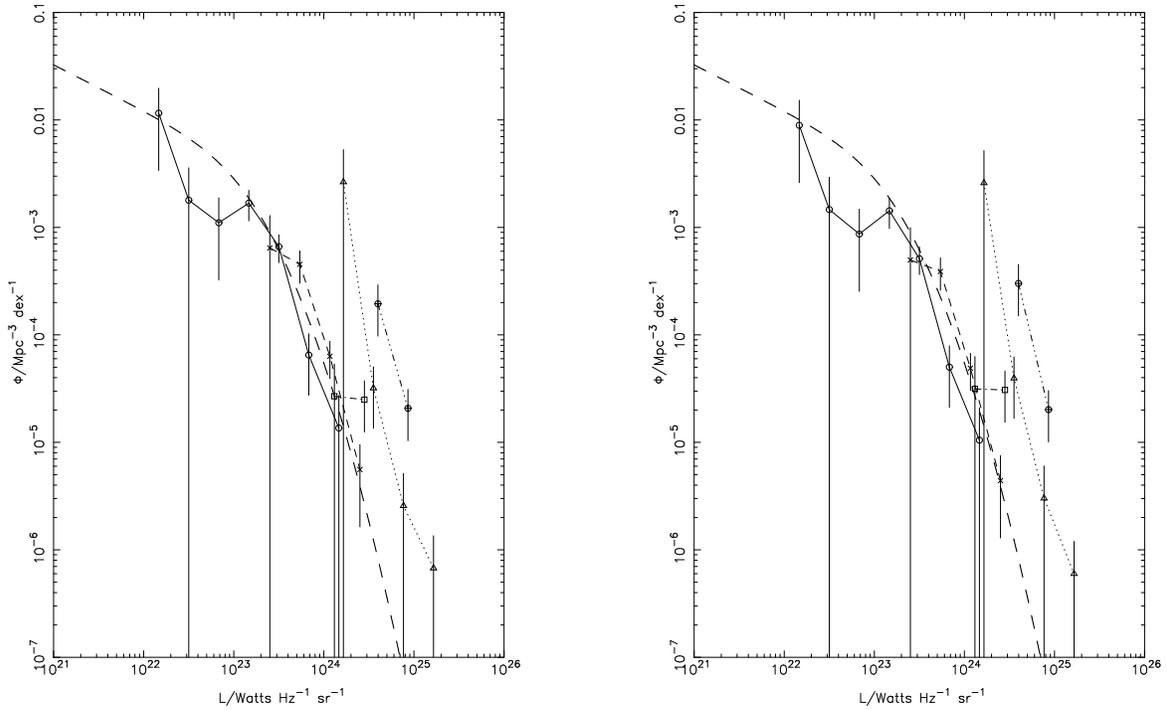}
\caption{Plots of the luminosity function in five redshift slices
at a rest-frame wavelength of 250 $\mu$m. In this figure no correction has
been made for flux-boosting.The left-hand panel shows
estimates of the luminosity function when no correction has been applied for
missing counterparts; the right-hand panel
shows the estimates after the correction
described in \S 4.4 has been applied. The key to the redshift 
slices is as follows:
open circles---$\rm 0 < z < 0.2$; crosses---$\rm 0.2 < z < 0.4$; squares---$\rm 0.4 < z < 0.6$;
triangles---$\rm 0.6 < z < 0.8$; crosses in 
circles---$\rm 0.8 < z < 1.0$. To guide the
eye in a rather complex diagram, thin lines link together the 
measurements in the
same redshift slice. The thick dashed line shows an estimate of
the local luminosity function at this wavelength by extrapolating in
wavelength from IRAS PSCZ survey using the information about the
spectral energy distributions of galaxies
from the SCUBA Local Universe and Galaxy Survey (Appendix 1).}
\end{figure*}

\begin{figure*}
\figurenum{8b}
\plotone{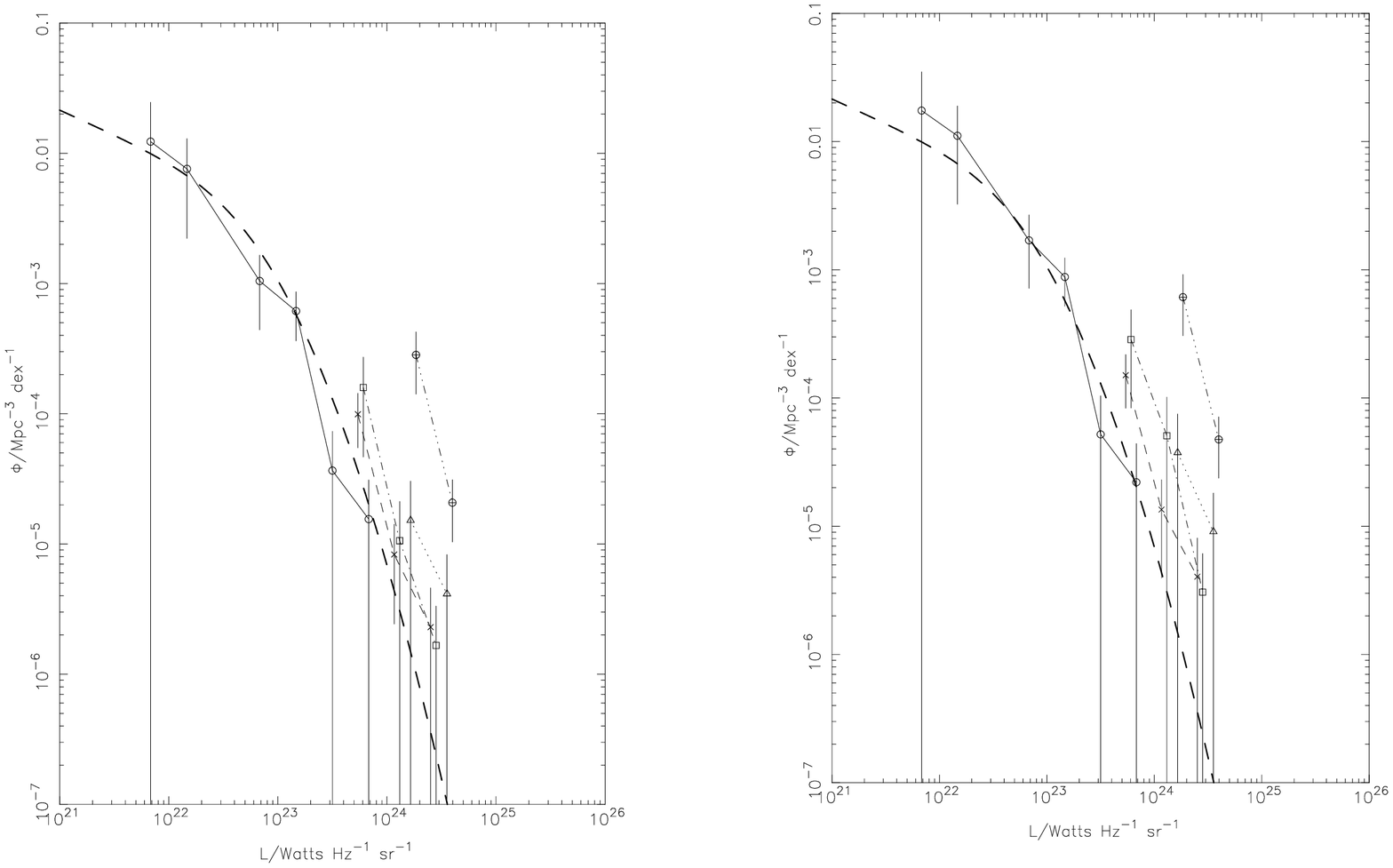}
\caption{The same as in (a) except at 350 $\mu$m.}
\end{figure*}

\begin{figure*}
\figurenum{8c}
\plotone{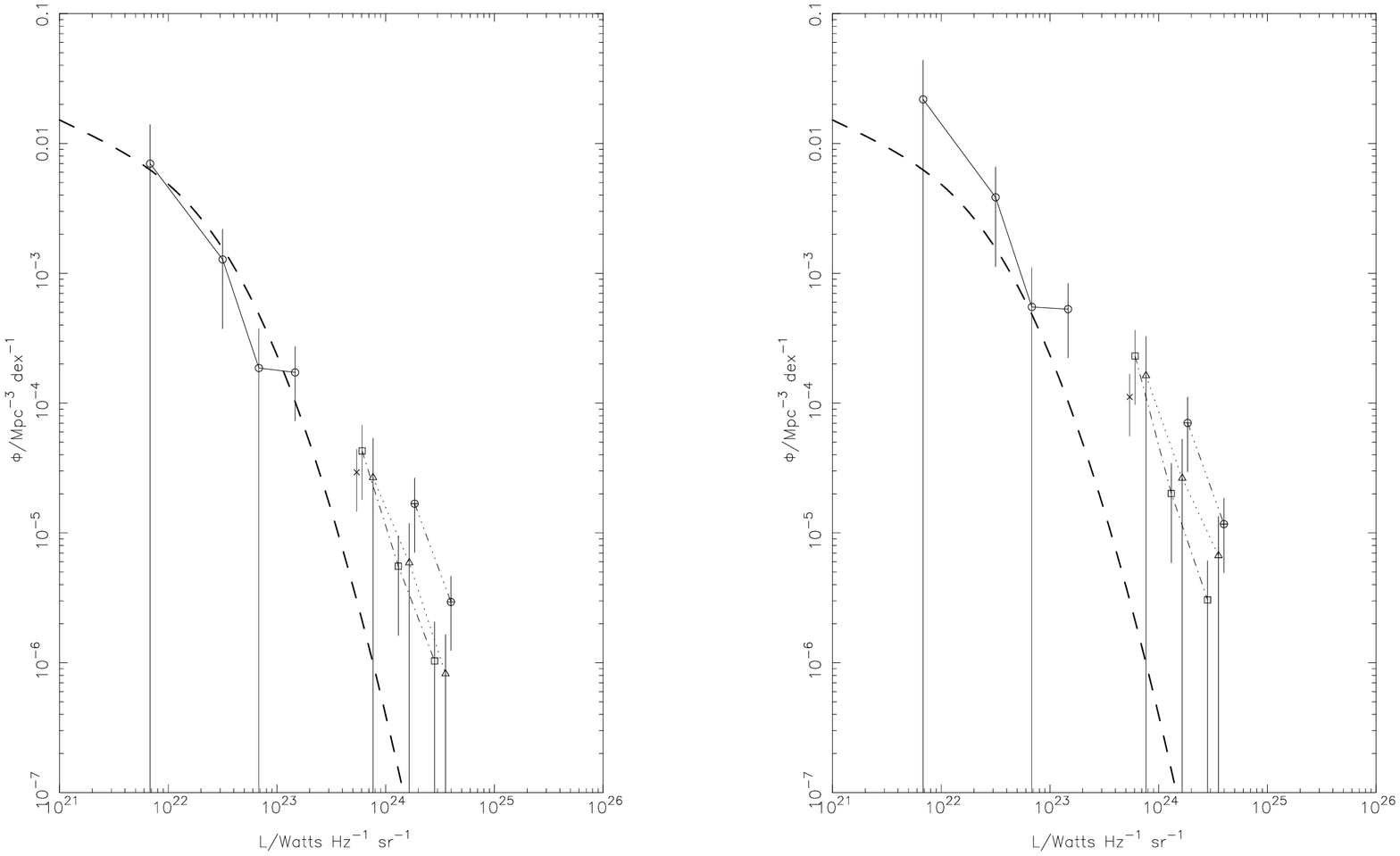}
\caption{The same as in (a) except at 500 $\mu$m.}
\end{figure*}

\begin{figure*}
\figurenum{9a}
\plotone{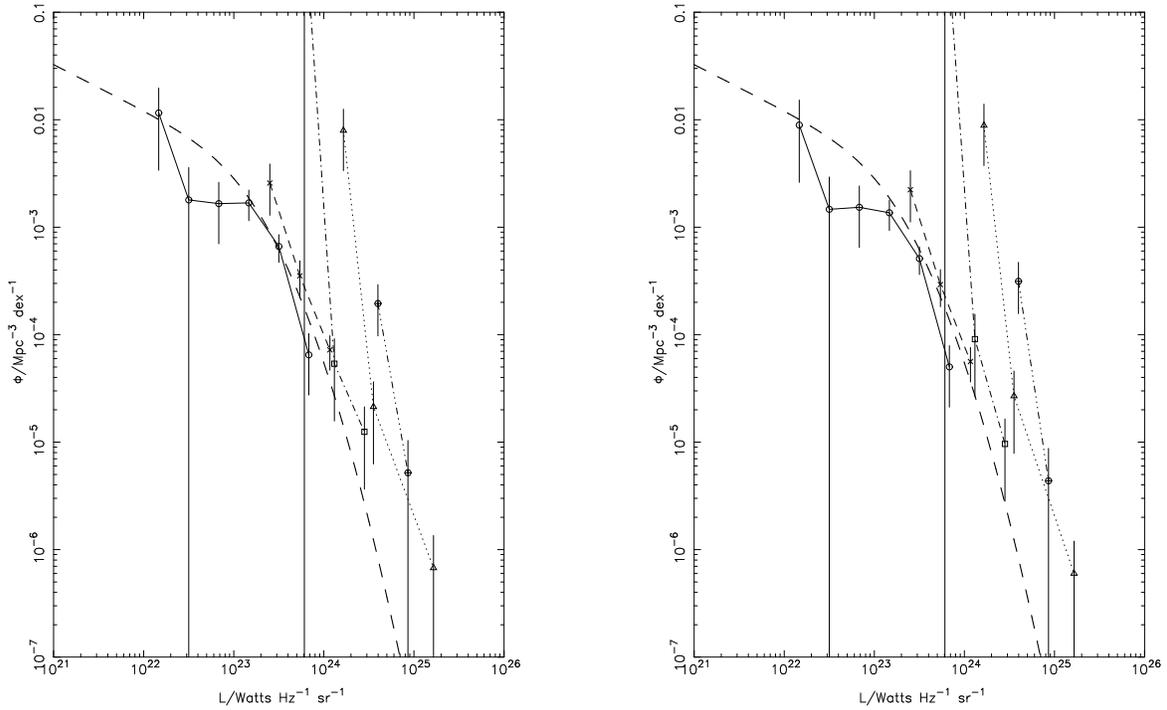}
\caption{Plots of the luminosity function in five redshift slices
at a rest-frame wavelength of 250 $\mu$m. In this figure a correction has
been made for
flux-boosting using the method described in Appendix 2.
The left-hand panel shows
estimates of the luminosity function when no correction has been applied for
missing counterparts; the right-hand panel
shows the estimates after the correction
described in \S 4.4 has been applied. The key to the redshift slices is as follows:
open circles---$\rm 0 < z < 0.2$; crosses---$\rm 0.2 < z < 0.4$; squares---$\rm 0.4 < z < 0.6$;
triangles---$\rm 0.6 < z < 0.8$; crosses in circles---$\rm 0.8 < z < 1.0$. To guide the
eye in a rather complex diagram, thin lines link together the measurements in the
same redshift slice. The thick dashed line shows an estimate of
the local luminosity function at this wavelength by extrapolating in
wavelength from IRAS PSCZ survey using the information about the
spectral energy distributions of galaxies
from the SCUBA Local Universe and Galaxy Survey (Appendix 1).}
\end{figure*}

\begin{figure*}
\figurenum{9b}
\plotone{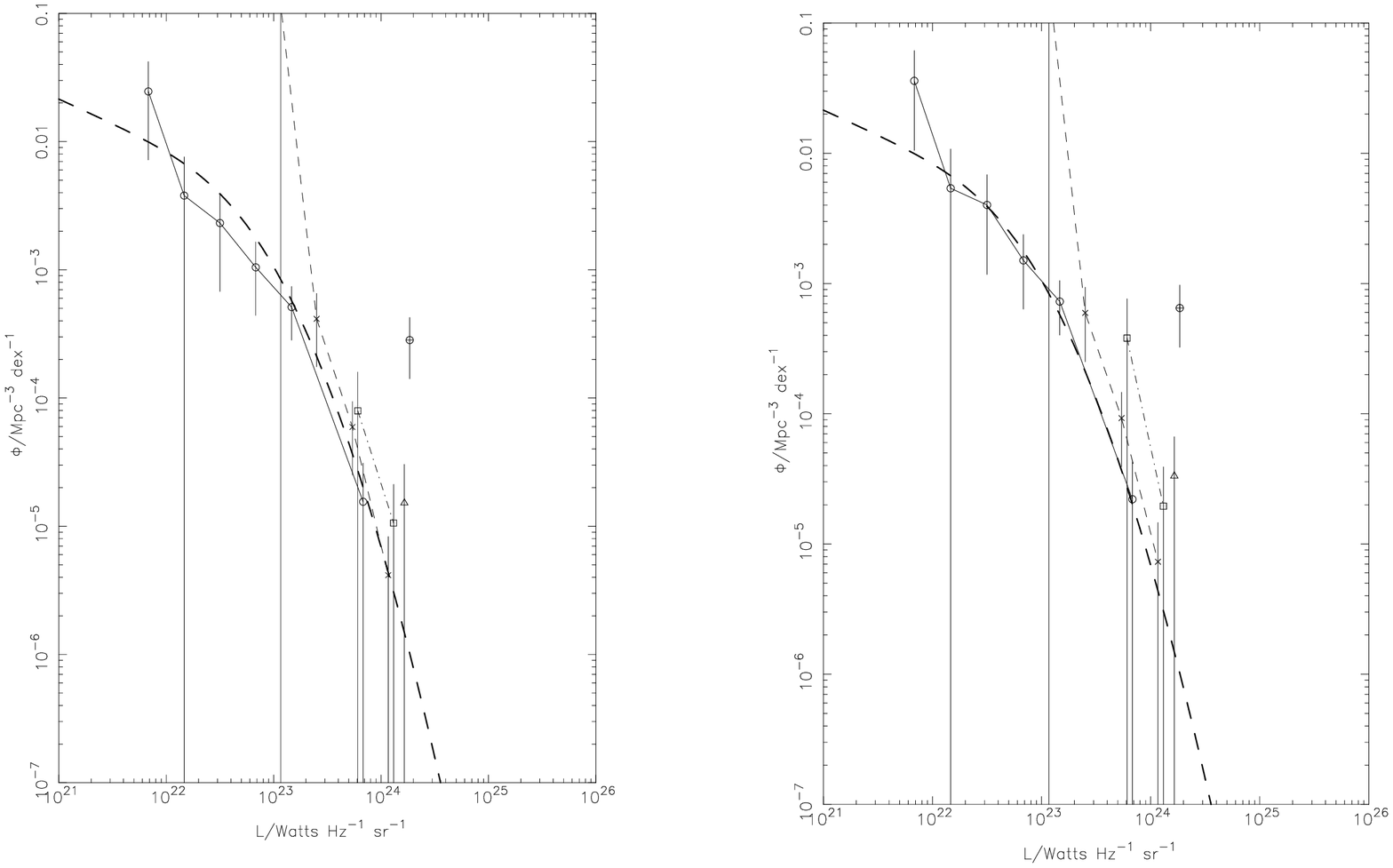}
\caption{The same as in (a) except at 350 $\mu$m.}
\end{figure*}

\begin{figure*}
\figurenum{9c}
\plotone{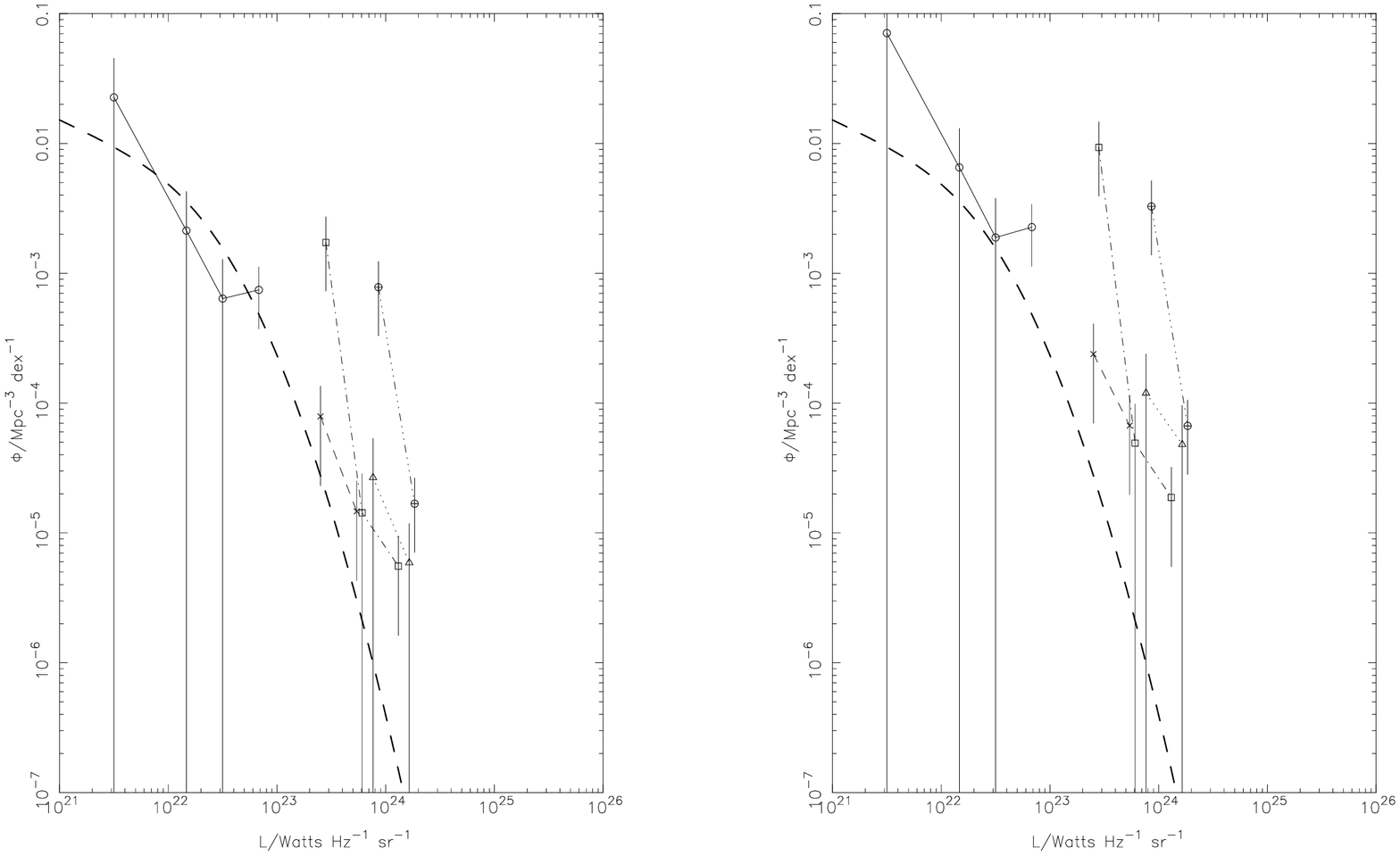}
\caption{The same as in (a) except at 500 $\mu$m.}
\end{figure*}

We chose to estimate the luminosity function in five redshift slices: $\rm 0 < z < 0.2$, $\rm 0.2 < z < 0.4$,
$\rm 0.4 < z < 0.6$, $\rm 0.6 < z < 0.8$, $\rm 0.8 < z < 1.0$.
To partially overcome the problem of the Eddington bias, rather than
using the flux densities in the original catalogue (D09), we remeasured the
fluxes from the BLAST images at the positions of the counterparts. 
To tackle the problem of the
missing counterparts, we made a minimal and maximal estimate of the luminosity function. We made the
minimal estimate by making no correction at all for the missing counterparts and the maximal
estimate by making the assumption that all the missing counterparts in Table 4 are actually
in the 24-$\mu$m/radio catalogues but have just been missed by our identification
technique. 
In calculating the average comoving accessible volume we assumed that all the sources
have the average spectral energy distribition found in D09.
In calculating the luminosities of the individual BLAST sources, we either
used the SED of the individual source given in D09 or, if that was
not possible, the average
SED.

We made our maximal estimate of the luminosity function by correcting for the
missing counterparts in the following way. We replace equation (7) by
\smallskip
$$
\phi \Delta log_{10} L = {\sum_{i=1}^n c_i \over V} \eqno(9)
$$
\smallskip
\noindent in which $c_i$ is the correction factor
for the i'th counterpart that falls in that particular
luminosity-redshift bin,
and the sum is over all the counterparts that fall in that bin.
We adopted the following scheme for estimating the correction factors.
We started by deriving the values using the same method that was used
to construct Fig. 7, except for the
three counterparts with very high values
of $c_i$. For these we used the average values of $c_i$ for the rest of the
counterparts at $\rm z > 0.2$. 
We then scaled all the values of $c_i$ by a constant factor so that the
number of missing counterparts predicted by equation (5) matched the
number of missing counterparts in Table 4. In doing this, we are
implicitly assuming that all the missing counterparts in Table 4
are actually in the 24-$\mu$m and radio catalogues but have
just been missed by our identification analysis. While this
is quite possibly true at 250 $\mu$m, it seems unlikely it is true
at 500 $\mu$m because of the much greater percentage of missing
counterparts. This assumption is why this method yields a maximal estimate
of the luminosity function.

The final question is how to deal with flux boosting. Although we have
remeasured the fluxes at the positions of the counterparts, our simulations (Appendix 2)
show that the fluxes are still too high. We dealt with this issue by
making estimates of the luminosity function both with and without making
a correction for flux-boosting. Appendix 2 describes how we have estimated
the effect of flux boosting in the BLAST images.

\clearpage
\begin{figure}
\figurenum{10}
\plotone{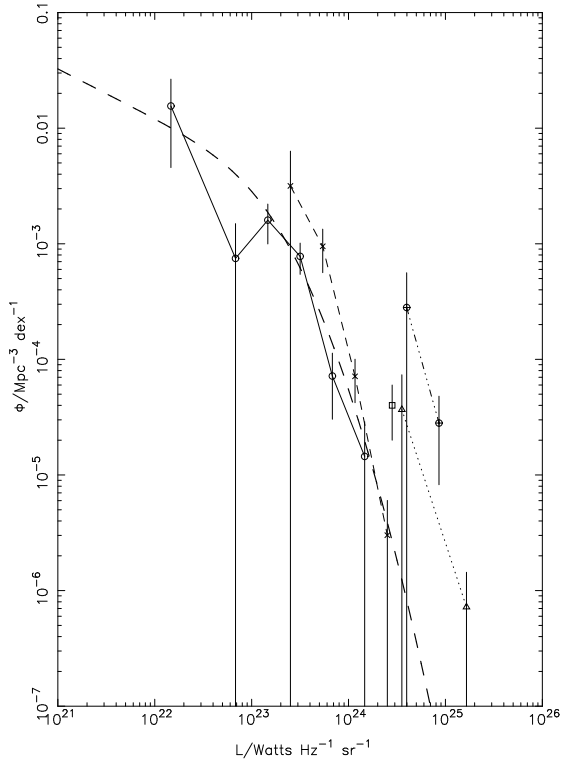}
\caption{Plots of the luminosity function in the five redshift slices
at a rest-frame wavelength of 250 $\mu$m, this time excluding
all galaxies in the FIDEL area. The symbols are the same as in
Figure 8. No correction has been made for flux boosting and
none has been made for the missing counterparts.}
\end{figure}

Figures 8 and 9 show our estimates of the luminosity functions 
at the three wavelengths in the five redshift slices.
Figure 8 shows our estimates of the luminosity function when no
correction is made for flux boosting. Figure 9 shows the effect of
including the correction for flux boosting.
On the left-hand side of each figure is the minimal estimate, without
any correction for the missing counterparts; on the right-hand side
is the maximal estimate.
These are the first measurements of the galaxy luminosity
function at these wavelengths. We have also plotted estimates of the
low-redshift luminosity function at these wavelengths, which have been obtained from the
IRAS PSCZ survey and the results of the only large submillimetre survey of nearby
galaxies (Appendix 1).

Inspection of the  figures shows that corrections for
flux boosting and the missing counterparts make very little obvious difference
to the estimates of the luminosity function.
The reason for this
is that the plots are on logarithmic axes and the luminosity bins
are very wide (0.33 in dex). This agreement gives us confidence that
the two obvious features of the luminosity functions are correct.
The first is that the agreement between the measurements of the 
luminosity function in the low-redshift
slice ($0 < z < 0.2$) and the extrapolation of the local luminosity function from
shorter wavelengths is surprisingly good. The second
is that there appears to be cosmic evolution, in the sense that
the space-density of the most luminous sources increases steadily with redshift.
The fact that there appears to be evidence for cosmic evolution in all
the sub-panels of Figures 8 and 9 suggests that this is a robust result.

As one additional check, we have calculating the 250-$\mu$m luminosity function using only
the BLAST sources outside the area covered by the FIDEL survey (\S 2). This 
tests whether the evolution could be the result of some peculiarity associated with
our deepest
optical/IR dataset, which also covers part of the
BLAST survey where the confusion is worst (\S 2). Figure 10, which 
contains no correction for flux boosting
or missing counterparts, shows that the evolution is still present even if we do
not use
the FIDEL dataset. 

This evolution in the space density of the most luminous sources has been seen before in the Spitzer bands \citep{huynh,floch},
but we might suspect that our results are adding something new because whereas the monochromatic luminosity
in the Spitzer bands is extremely sensitive to the temperature of the dust, in the BLAST bands the monochromatic
luminosity is equally sensitive to the mass of dust that is present. Thus Fig. 8 suggests that there may be
strong cosmic evolution not only in the luminosities of galaxies but in the masses of dust in the galaxies.
We can test whether this is so by calculating the space-density of galaxies as a function of dust mass. We can
do this using a straightforward adaptation of the formalism above.
The monochromatic luminosity at a frequency $\nu$ is connected to the mass of dust in a galaxy by
the relation:
\smallskip
$$
L_{\nu} = B_{\nu}(T_d) \kappa_{\nu} M_d  \eqno(10)
$$
\smallskip
\noindent in which $B_{\nu}$ is the Planck function, $T_d$ and $M_d$ are the dust
temperature and dust mass, and 
$\kappa_{\nu}$ is the dust-mass opacity coefficient. 
Although galaxies clearly contain dust with a range of dust temperatures, Dunne
and Eales (2001) have shown that most of the dust, even for a ULIRG like Arp
220, has a temperature of only $\simeq$ 20 K. In using equation 10 to make
the connection between dust mass and luminosity we have assumed a dust temperature
of 20 K and the value of the
dust-mass opacity coefficient from James et al. (2002), extrapolating this to the
BLAST frequencies assuming that it scales as $\nu^2$. We have used the
sample of galaxies at 250 $\mu$m (Table 4) because the
percentage of sources with counterparts is highest at this wavelength. 
The dust-mass function (the space-density of galaxies as a function of dust mass)
in a bin in the mass-redshift plane is then given by
\smallskip
$$
\phi(M_1 < M_d < M_2,z_1<z<z_2) \Delta log_{10} M_d = {n \over V} \eqno(11)
$$
\smallskip
\noindent in which $n$ is the number of galaxies with dust masses
and redshifts that fall within this bin and $V$ is calculated using
equation (8). Figure 11 shows the results for the five redshift
slices without making any correction for missing counterparts.
There is clearly strong evolution, in the sense that the space-density
of the galaxies with the highest dust masses increases steadily with redshift.
Pascale et al. (2009) concluded that there was no evolution in the
comoving density of dust in the universe. However, Pascale et al.
effectively
measured $\phi <M_d>$ in each redshift slice, and Figure 11 shows
that this does not change very much: the average dust mass of the
galaxies detected at low redshift is lower than at high redshift
but their space-density 
is higher. It is only by comparing the space-density
at different redshifts but at the same dust mass
that it is possible to see the evolution.

\begin{figure}
\figurenum{11}
\epsscale{0.8}
\plotone{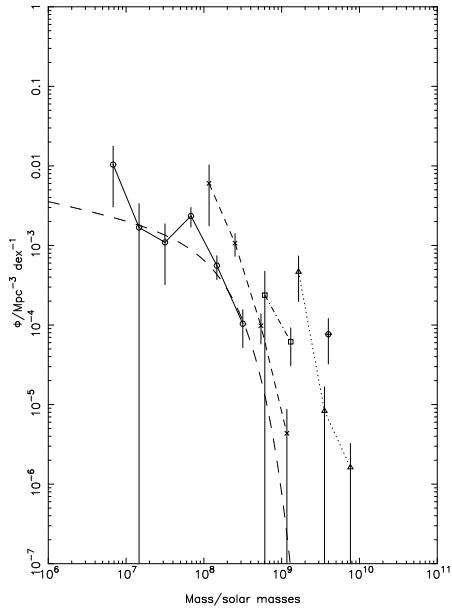}
\caption{Plot of the `dust-mass function', the space density of galaxies
as a function of dust mass. We have estimated this for five redshift slices,
and the key is the same as for Fig. 8.
The thick dashed line shows the Schechter function that is the best fit to
the dust-mass function in the lowest redshift slice.}

\end{figure}

Note that the result in Fig. 11 
is very insensitive to our assumptions about temperature,
because on the Rayleigh-Jeans tail the Planck function in equation 10
only depends on the first power of dust temperature. The strength of the
evolution would be less if the temperature of the bulk of the
dust at high redshift were higher than that at low redshift. But even if the
temperature were a factor of two greater at high redshift, the effect on the
high-redshift points in Figure 11 would be to move them a factor of
two to the left, which is not enough to remove the result. 
It is possible to think of scenarios in which the evolution was
caused by temperature. Suppose that as one moves to higher redshift, the
fraction of BLAST galaxies that contain a luminous but obscured
quasar gradually increases, and by a redshift of $\sim 1$ the temperature
of the dust heated by the hidden quasar
is a factor of 10 greater than at $z=0$. This would explain
the evolution seen in Fig. 11. However,
because of the strong dependence of bolometric luminosity on temperature,
this increase in temperature would correspond to a increase in
bolometric luminosity of at least a factor of $10^5$.
Therefore, it is much 
harder to explain the evolution visible in Fig. 11 as a temperature effect
than as an increase in the number of galaxies with high dust masses.

\section{Concluding Remarks}

We have carried out a redshift survey of the sources found in the BLAST
survey of GOODS-South. Our basic results are as follows:

\begin{itemize}

\item The equivalent widths of the H$\alpha$ line show that the counterparts
to the BLAST sources are mostly star-forming galaxies with a mean equivalent width
similar to that for the
star-forming galaxies found in 
the Sloan Digital Sky Survey and the 2dF Galaxy
Redshift Survey. Therefore, the BLAST counterparts appear to be star-forming
galaxies but not particularly extreme ones.

\item Approximately one quarter of the BLAST counterparts contain an active nucleus, judged
either by the line ratios or the presence of broad emission lines.

\item We have made an unbiased estimate of the errors in the redshifts
produced by the photometric redshift methods developed from the
COMBO-17 survey \citep{wolf} and the SWIRE survey (Rowan-Robinson et al.
2008). 
Using $\delta = { z_{phot} - z_{spec} \over 1 + z_{spec} }$
as our measure of the discrepancy between the photometric and spectroscopic redshift,
we found that 8\% of COMBO-17 photometric redshifts and 9\% of SWIRE
redshifts had catastrophic errors in the sense that $|\delta| > 0.15$.
Exluding these catastrophic errors, we found that errors ($\sqrt{<\delta^2>}$)
were 0.031 for the COMBO-17 redshifts and 0.056 for the SWIRE redshifts.

\item We have used the redshifts and the images to investigate the
30\% of BLAST sources that have two or more counterparts. We conclude that
there is evidence in at least half the cases that the two
counterparts are physically associated, either
because they are interacting or because 
they are in the same large-scale structure. 

\item We have made the first estimates of the luminosity function at
the three BLAST wavelengths and in five redshift slices. We find strong evolution,
in the sense that the space-density of the most luminous sources increases
steadily with redshift out to $\rm z = 1$.

\item We have also investigated the evolution of the dust-mass
function with redshift, finding gradual evolution in the space-density
of the galaxies with the highest dust masses out to $\rm z = 1$.

\end{itemize}

The most interesting result is probably the last one. It is well known that
the luminosity-density of the universe evolves strongly with redshift, whether
observed in the optical waveband or the far-IR/submillimetre wavebands, and that
the space-density of the most luminous sources evolves strongly with redshift \citep{simon,huynh,floch,mag}.
But this increased luminosity-density need not necessarily be associated with
an increase in the amount of interstellar matter in galaxies. 
If the increased luminosity-density is caused by an increase in the
global star-formation rate, it is possible, for example, that this is caused
by a larger number of galaxy interactions at high redshift, which trigger starbursts, and
not necessarily by the larger amount of interstellar material in galaxies.
However, Figure 11 
shows that the space-density of galaxies with high dust masses, and thus presumably large
reservoirs of interstellar material, is also evolving strongly with redshift.


\acknowledgments{We are grateful to Heath Jones for his help
with the observations and Rob Sharp for his help with the 2dfdr
data-reduction pipeline.
This work makes use of the Runz redshifting code developed by Will Sutherland,
Will Saunders, Russell Cannon and Scott Croom, and we are grateful to
Scott Croom for making this available to us.
We thank Seb Oliver for making available the SWIRE optical images
and for a useful conversation about the SWIRE photometric redshift
method and Luca Cortese for emergency help with iraf.
We acknowledge the support of NASA through grant numbers NAG5-12785, NAG5-13301
and NNGO-6GI11G, the NSF Office of Polar Programs, the Canadian Space Agency, the
National Sciences and Engineering Research Council of Canada and the UK Science
and Technology Facilities Council. This paper relies on observations
made with the AAOmega spectrograph on the Anglo-Australian Telescope, and we
thank the staff of the telescope and especially those involved in the devlopment of
the spectrograph.

}

\appendix

\section{Extrapolating the Local Luminosity Function from IRAS Measurements}

The local luminosity function is well known in the wavelength range
$10 \leq \lambda < 100 \mu m$ because of the all-sky IRAS survey.
However, the problem in simply estimating the local luminosity
function at the BLAST wavelengths by extrapolating an IRAS luminosity function
is that it is not obvious what spectral energy distribution to use;
the galaxies
we know most about are those that were detected by IRAS but these 
are likely to contain warmer dust than
the galaxies detected at longer wavelengths in the BLAST survey.
Serjeant and Harrison (2005) suggested a way of overcoming this problem
by using the results of the largest submillimetre survey of nearby galaxies:
the SCUBA Local Universe and Galaxies Survey (SLUGS) \citep{dunne2000,dunne2001,cat2005}.
This is the method we have adopted here.

We started with the $\sim$10,000 galaxies detected in the IRAS PSCZ survey \citep{will},
which was a survey of 82\% of the sky down to a flux density of $S_{60 \mu m} \succeq 0.6$ Jy.
We only included in our analysis galaxies that had both 60 and 100 $\mu m$ detections
and, to avoid the effects of peculiar motions and evolution, velocities
between 300 and 30,000 km s$^{-1}$. We calculated the accessible volume for
each galaxy using both the flux and the velocity limits. We then used
the tight relation that exists between the ratio of 60 to 100 $\mu$m flux
and the ratio of 60 to 850 $\mu$m flux that was discovered in the
SLUGS survey \citep{dunne2001} to estimate the flux of each galaxy at
850 $\mu$m. The precise form of the relationship we used is the one
given by Vlahakis et al. (2005):
\smallskip
$$
log_{10} {S_{60} \over S_{100}} = 0.365 log_{10} {S_{60} \over S_{850}} - 0.881 \eqno(A1)
$$
\smallskip
We fitted a simple two-component dust model to the 60, 100 and 850 $\mu$m
values, which allowed us to estimate the luminosity of each galaxy at
the BLAST wavelengths. We then used equation 6 to estimate the
local luminosity function at the three BLAST wavelengths. Finally,
we fitted the modified Schechter function that Saunders et al.
(1990; equation 6.1) found was a good fit to the
60-$\mu$m luminosity function to the three BLAST luminosity functions.
This is the function plotted in Figs 8-10.

\begin{figure*}
\figurenum{A1}\epsscale{0.8}
\plotone{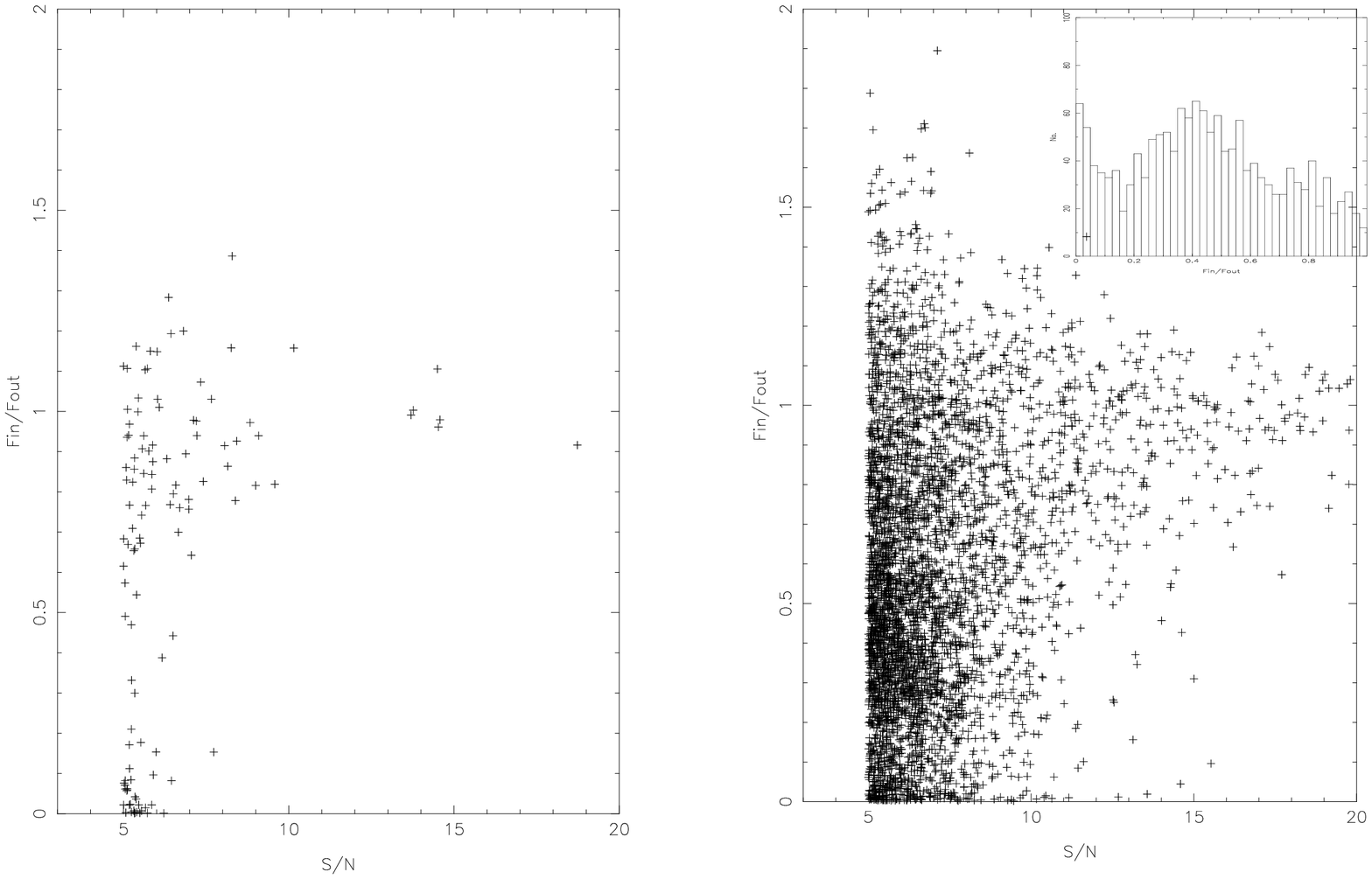}\caption{Plot of the ratio of the input (true) flux to the measured flux for
the brightest components of the sources in our artificial 5$\sigma$ catalogue.
The left-hand figure is for our simulation of the wide survey, the
right-hand figure for our simulation of the deep survey. The inset to
the right-hand figure shows the histogram of this flux ratio for
the sources between 5 and 6$\sigma$ in the deep catalogue. This
shows more clearly than the main figure that there is a peak at
${F_{true} \over F_{meas}} \simeq 0$,
showing there are probably spurious sources in both parts of the BLAST survey.}
\end{figure*}

\section{Modelling a 5$\sigma$ Catalogue}

Dye et al. (2009) presented a 5$\sigma$ catalogue of BLAST sources, which
we have also used in this paper. This catalogue is affected by Eddington
bias (\S 4.4), and we have used a Monte-Carlo simulation of the BLAST fields
to investigate the effects of Eddington noise on the catalogue, and
in particular to investigate the influence of these effects on our estimates
of
the luminosity function described in \S 4.

We used the source counts from Patachon et al. (2009) and our maps of instrumental
noise to generate Monte-Carlo realizations of the deep BLAST image and of the
shallower wide image. We have not incorporated any clustering of the BLAST
sources, since Patanchon et al. (2009) concluded that clustering has only a small
effect in models of the effect of Eddington bias.
After generating the maps, we used the source-finding software used on the
real BLAST images to generate a catalogue of 5$\sigma$ sources. 
We call these sources the `output sources'.
For each of these sources
we then found all the sources that were used to create the realization within
0.5$\times$FWHM (full-width half maximum of the telescope beam) of the 
position of the output source; we call these sources the `input sources'.

In estimating the luminosity function (\S 4.4), we used fluxes measured from
the BLAST maps at the positions of the counterparts. We
modelled this procedure by making the assumption that the brightest
of the input sources represents the submillimetre emission associated
with the counterpart. We remeasured the submllimetre fluxes from the
artificial submillimetre maps at the position of the brightest input source. These
fluxes, which we call $F_{out}$, thus represent the fluxes we used
to estimate the luminosity functions shown in Figure 8. We call the true
fluxes of these sources $F_{in}$. Figure A.1 shows the ratio of $F_{in}$ to
$F_{out}$ plotted against the signal-to-noise of the output source
for both the wide and the deep images. 

These two figures show two
effects of Eddington bias. First, the panels for the wide and deep surveys
suggest that some of the 5$\sigma$ sources in both regions are either
instrumental noise, promoted by Eddington bias to appear as actual sources, or
a confused combination of instrumental noise and many faint sources. This is
most apparent in the figure for the wide survey, which is dominated by
instrumental rather than confusion noise. This figure shows a cluster
of sources with ${F_{in} \over F_{out}} \simeq 0$, which 
can clearly not be associated with a single luminous source.
This feature can also be seen 
in the figure for the deep survey (see the inset to the right-hand figure).
An alternative empirical way to determine the fraction of spurious sources is,
of course, to look for optical counterparts: the number that do not have counterparts
gives an upper limit to the percentage of sources that are effectively
instrumental noise (Table 4). 
The results of this simulation are why we suggest in \S 4.4 that some of
the sources with missing counterparts may actually not be genuine sources.
If this is true, the luminosity functions on the left-hand sides of
Figures 8 and 9 are likely to the best estimates of the genuine
luminosity functions.

\begin{figure}
\figurenum{A2}
\epsscale{0.8}
\plotone{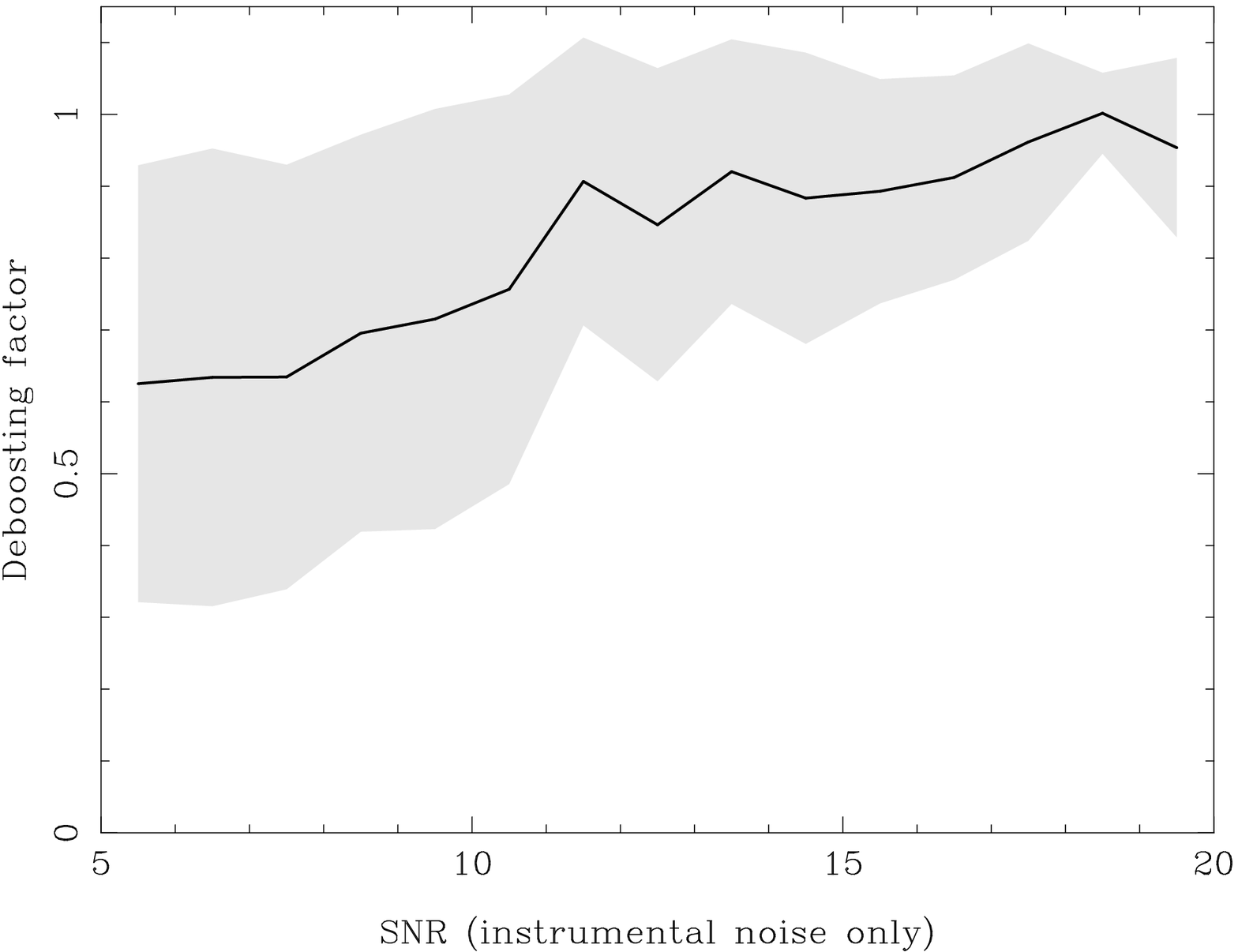}
\caption{Plot of the deboosting factor at 250 $\mu$m for the deep part
of the BLAST survey used to correct the luminosity function in Figure 9.
This is the average value of ${F_{true} \over F_{measured}}$ for the
brightest component of the sources in the artifical catalogue. The
shaded area shows the standard-deviation of this factor, showing
that the deboosting factor for individual sources is highly uncertain.}
\end{figure}

The second effect is the bias on the fluxes measured at the counterpart.
Although we adopted this procedure to mitigate the effect of
flux boosting, flux boosting can still clearly be a big effect. To quantify this effect,
we have measured the average value of $F_{in} \over F_{out}$ as a function of
the signal-to-noise of the output source. We excluded all sources
with ${F_{in} \over F_{out}} < 0.2$, which we argue are either
instrumental noise or that do not represent single luminous sources. 
Fig A.2 shows the results at 250-$\mu$m for the deep
survey. We have used tables constructed from figures like this
to `deboost' the fluxes used to estimate the luminosity functions
shown in Figure 9.

There is also a third effect, which is that an output source is composed
of more than one input source. Figure A.3 shows a histogram of the
ratio of the brightness of the second brightest input source to the brightest
input source. The figure suggests that most output sources are dominated by
a single input source, although 21\% of the sources in the artifical catalogue made
for the deep survey have a second input source
that is over 50\% of the brightness of the brightest input source.
We have made no correction for this effect, although a simple thought
experiment suggests that this effect effectively
operates in the opposite direction to the flux boosting effect.
Suppose an output source is composed of three input sources of equal
brightness. If we make the assumption that these sources
also all have the same redshift, the correction we should make to
a point on the luminosity function is to move it to
a luminosity that
is three times lower and to a number-density that is three times
higher. Since this correction is roughly parallel to a typical
luminosity function, the net effect is relatively small. Therefore,
by only correcting for the flux-boosting effect in Figure 8, we
are essentially putting an upper limit on the effect of Eddington
bias on the luminosity function.

\begin{figure*}
\figurenum{A3}
\epsscale{0.8}
\plotone{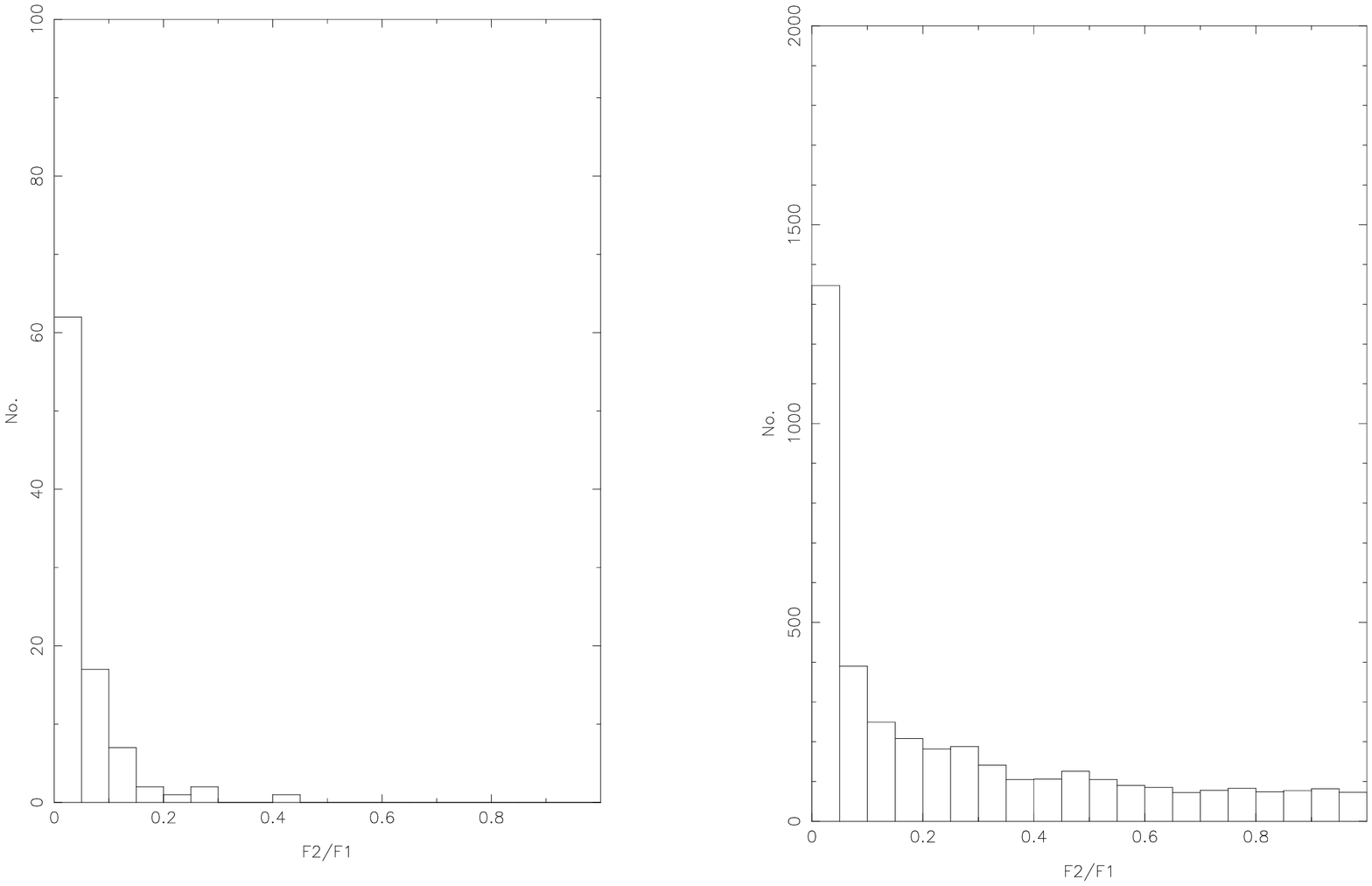}
\caption{Ratio of the input (true) flux of the second brightest
component of each source in the artificial catalogue to the
input (true) flux of the brightest component. The figure on the
left is for the wide survey, on the right for the deep survey.}
\end{figure*}

\clearpage



\begin{figure}
\figurenum{1}
\epsscale{.80}
\plotone{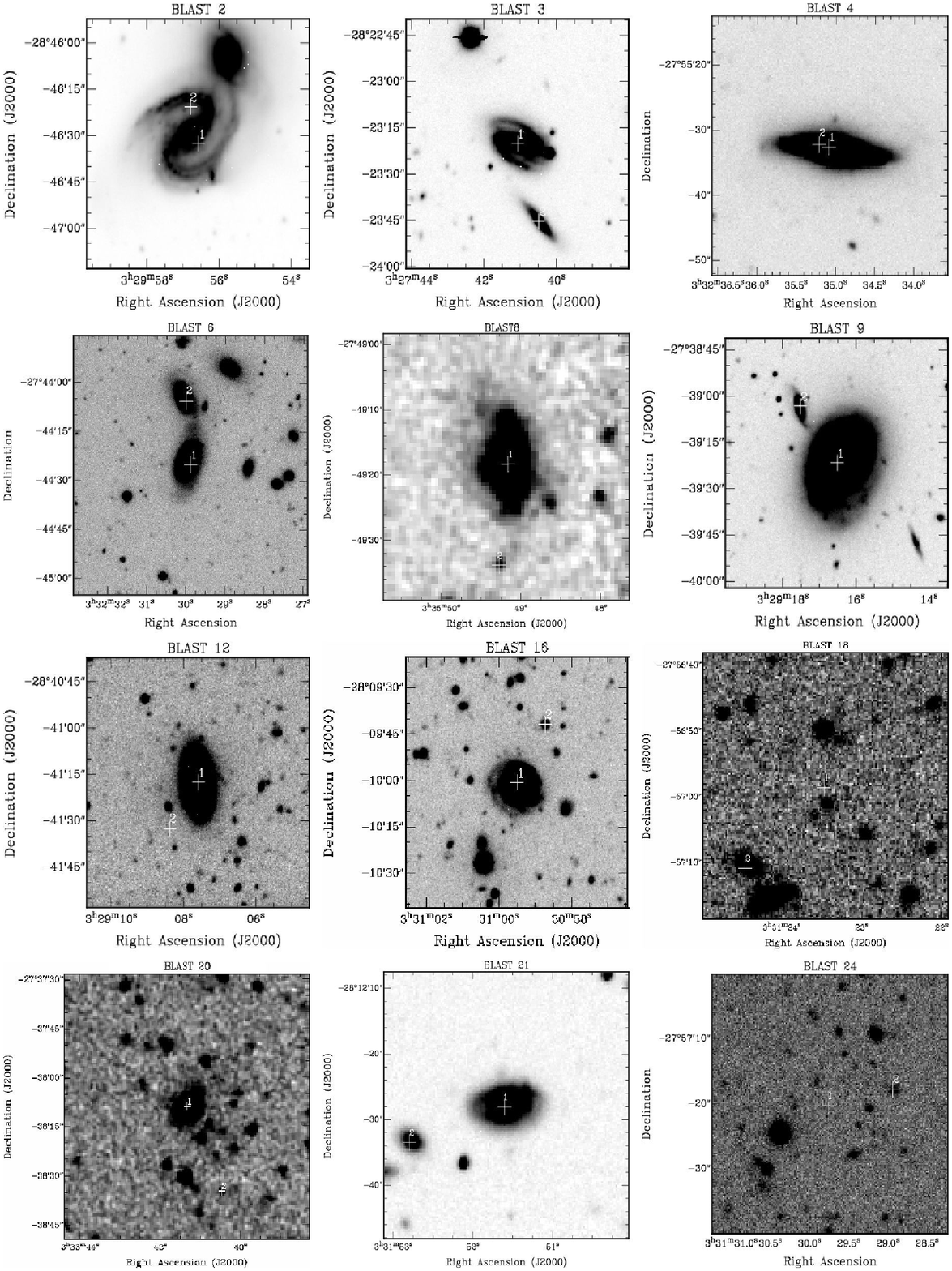}
\caption{Images of the BLAST sources for which there are at least two counterparts.
The positions of both counterparts are marked by crosses. The images
are of a field $\rm 40 \times 40\ arcsec^2$ in size for all sources except
for the following sources for which the fields are $\rm 80 \times 80\ arcsec^2$
in size: 2,3,6,9,12,16,20,32,35,44,53,57,70,73,76,80,95,
96,113,115,120,152,173,196,197,253 and 320. 
The images are taken from the COMBO-17 R-band image
for the sources 4,6,24,26,35,55,131,162 and 265; 
from the IRAC 3.5 $\mu$m image
for the sources 8,20,32,37,39,53,57,64,76,77,93,
95,96,103,106,113,
118,120,123,139,152,175,204,205,253,257,
320; and from the SWIRE r-band image
for the remaining sources.}
\end{figure}

\clearpage
\begin{figure}
\figurenum{1(b)}
\epsscale{1.0}
\plotone{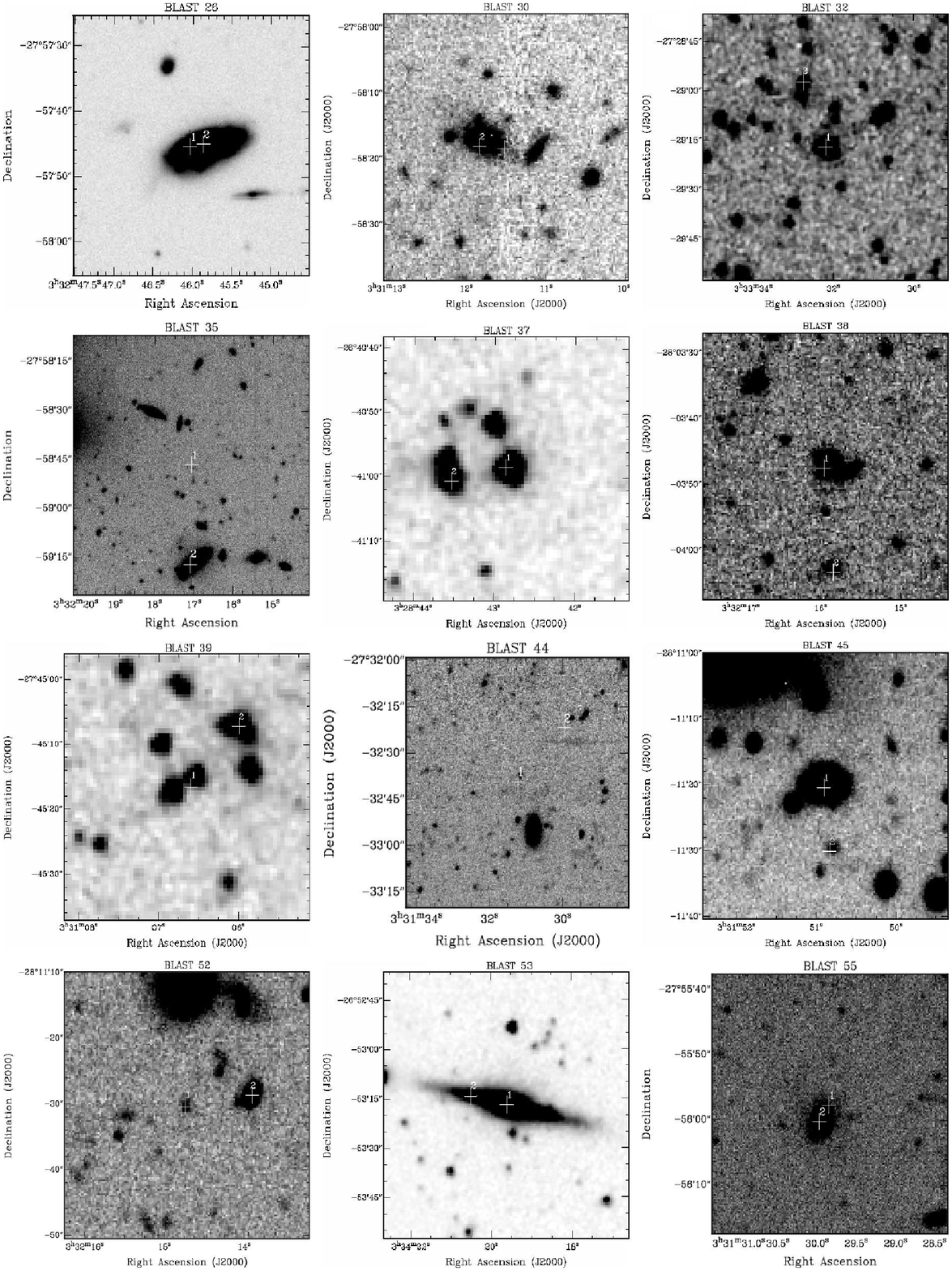}
\caption{Figure 1, continued}
\end{figure}
\clearpage

\clearpage
\begin{figure}
\figurenum{1(c)}
\plotone{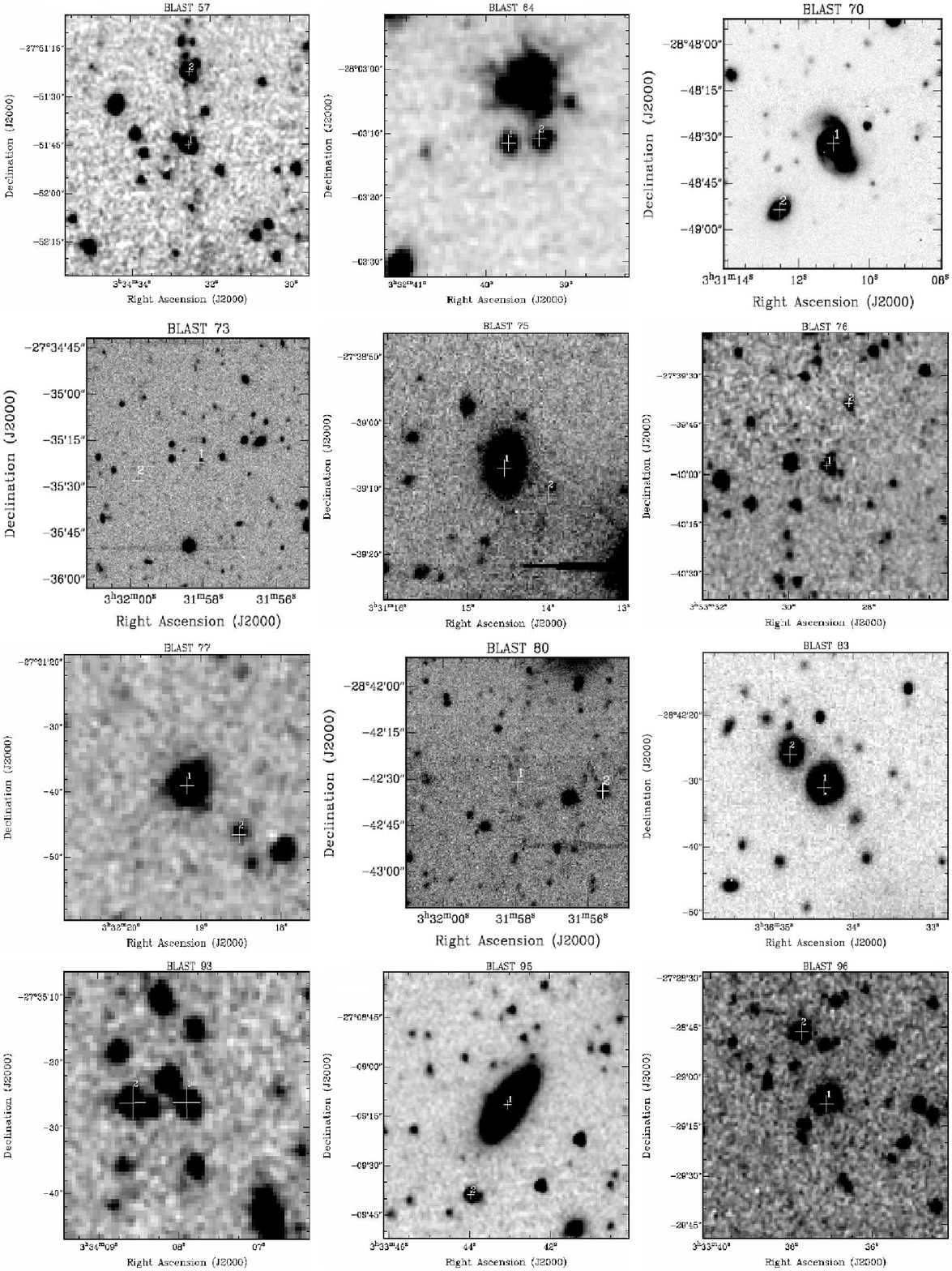}
\caption{Figure 1, continued}
\end{figure}
\clearpage

\clearpage
\begin{figure}
\figurenum{1(d)}
\plotone{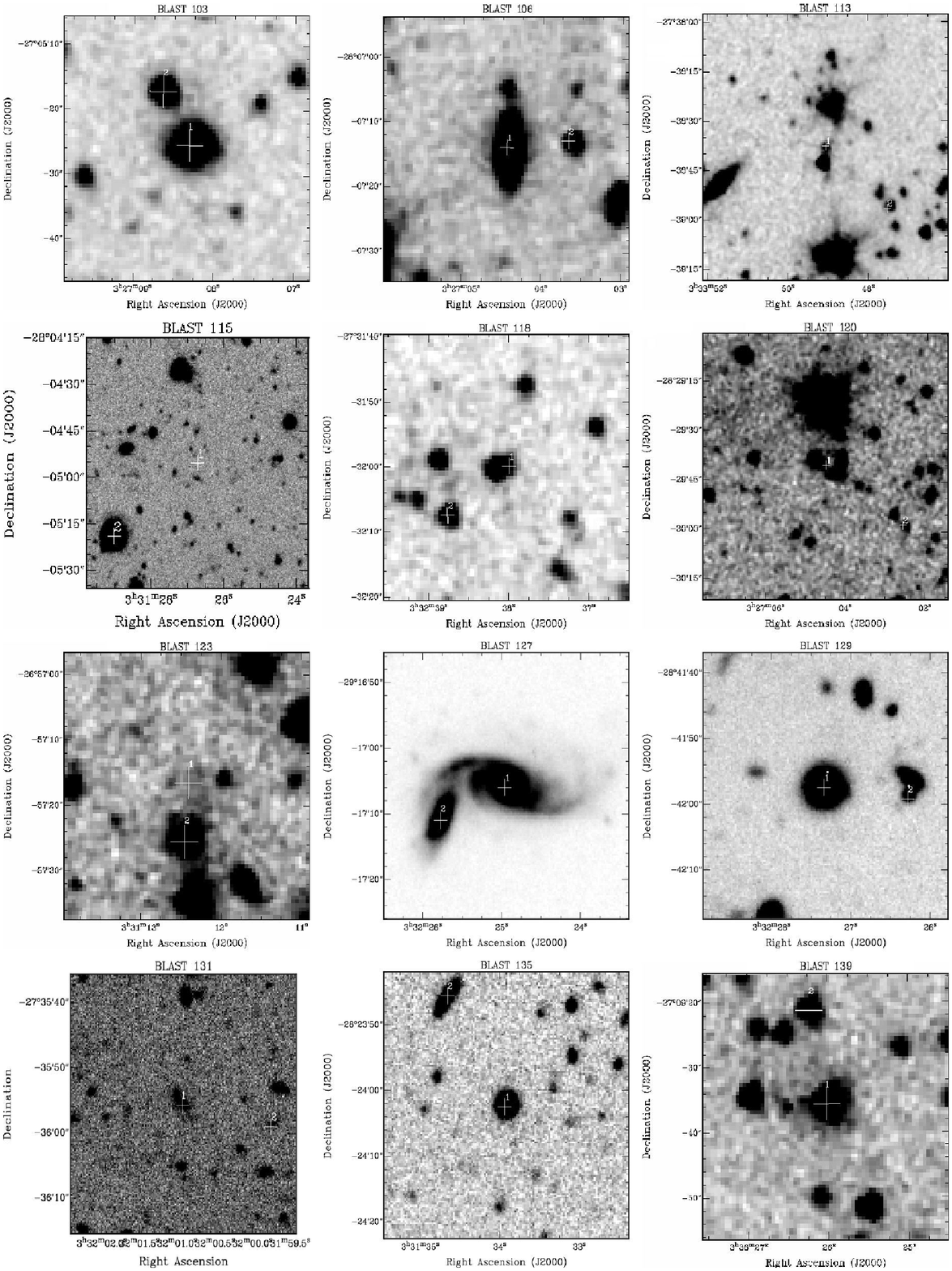}
\caption{Figure 1, continued}
\end{figure}
\clearpage

\begin{figure}
\figurenum{1(e)}
\plotone{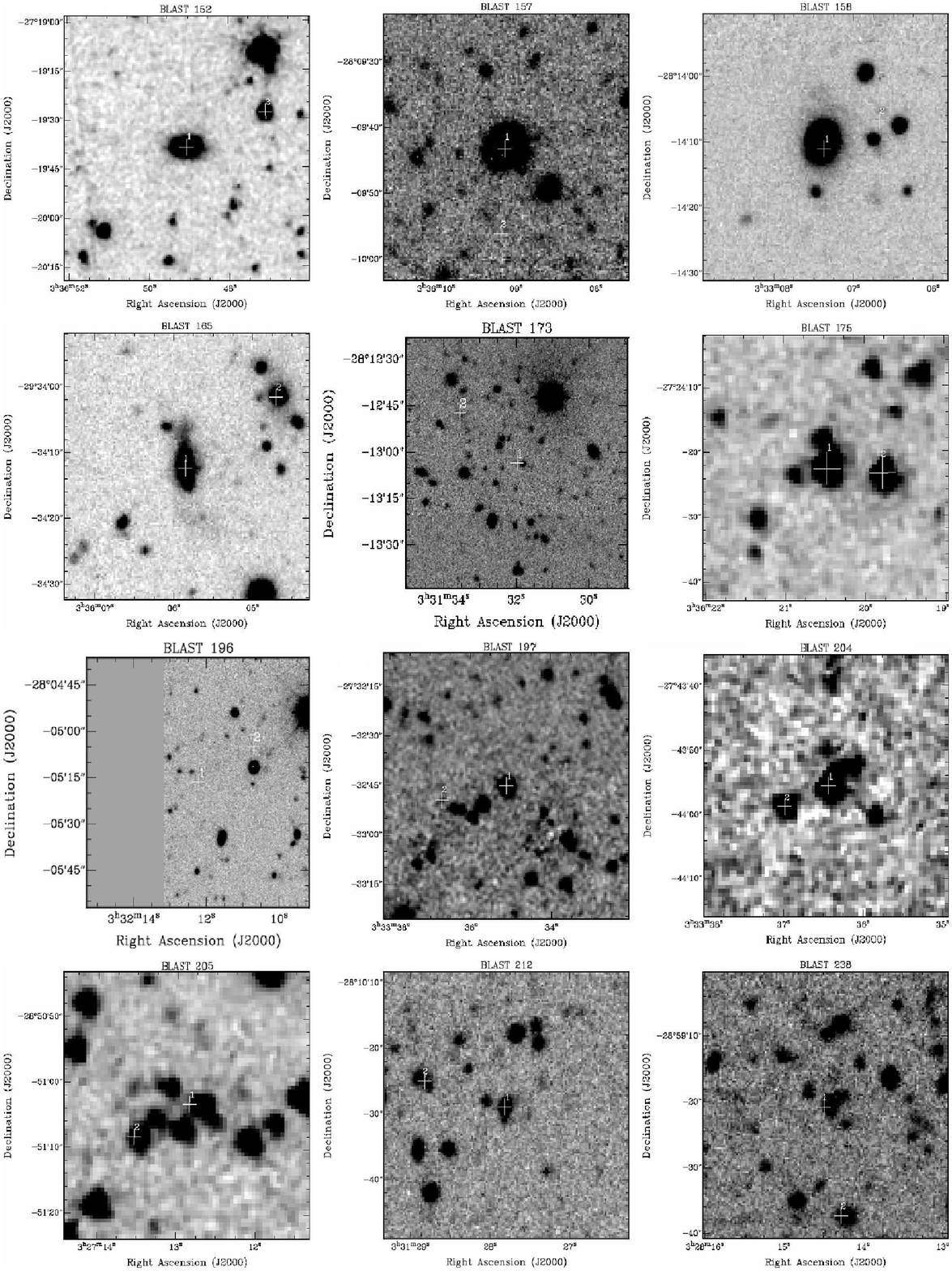}
\caption{Figure 1, continued}
\end{figure}
\clearpage

\begin{figure}
\figurenum{1(f)}
\plotone{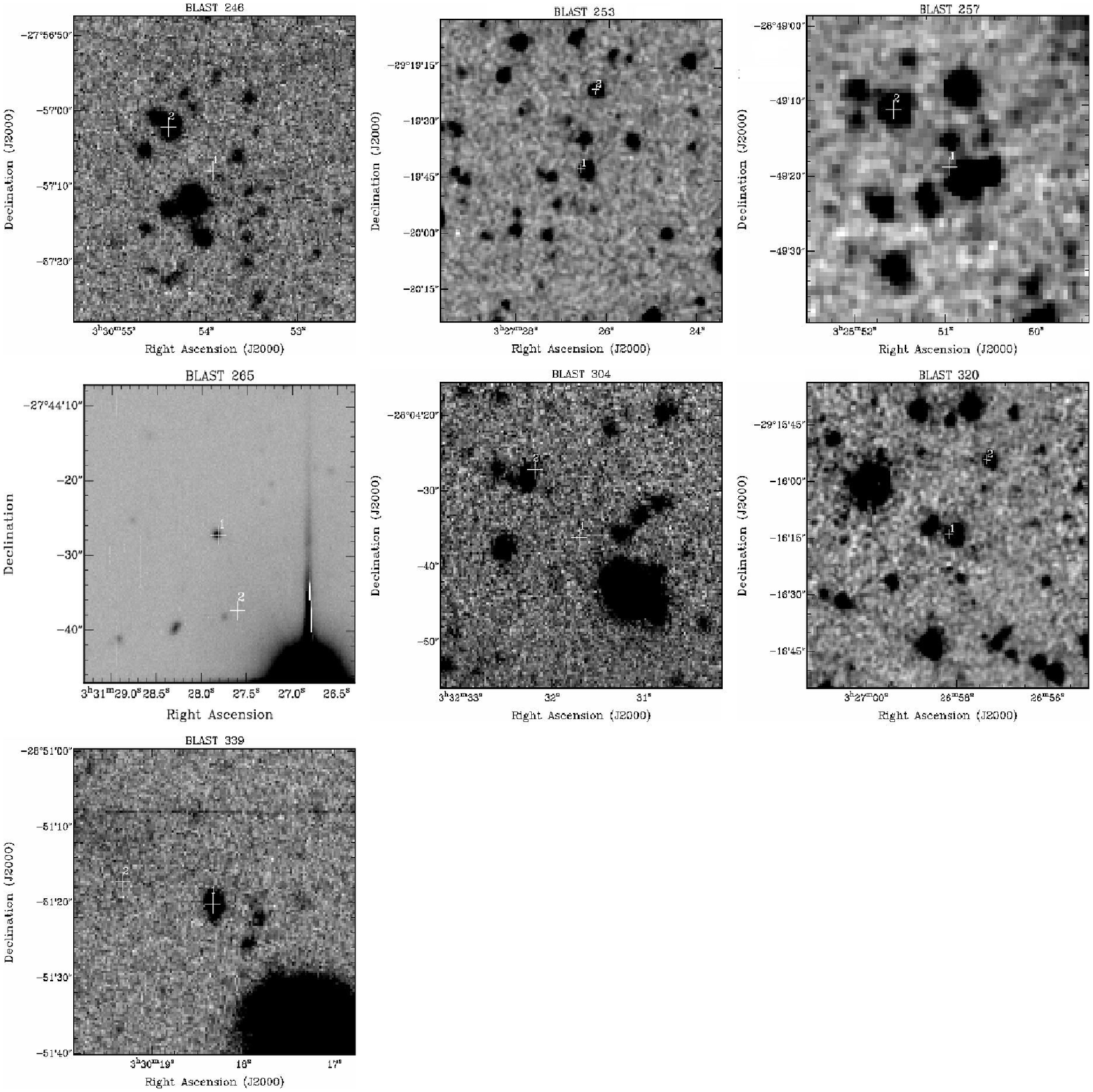}
\caption{Figure 1, continued}
\end{figure}
\clearpage

\begin{figure}
\figurenum{2a}
\plotone{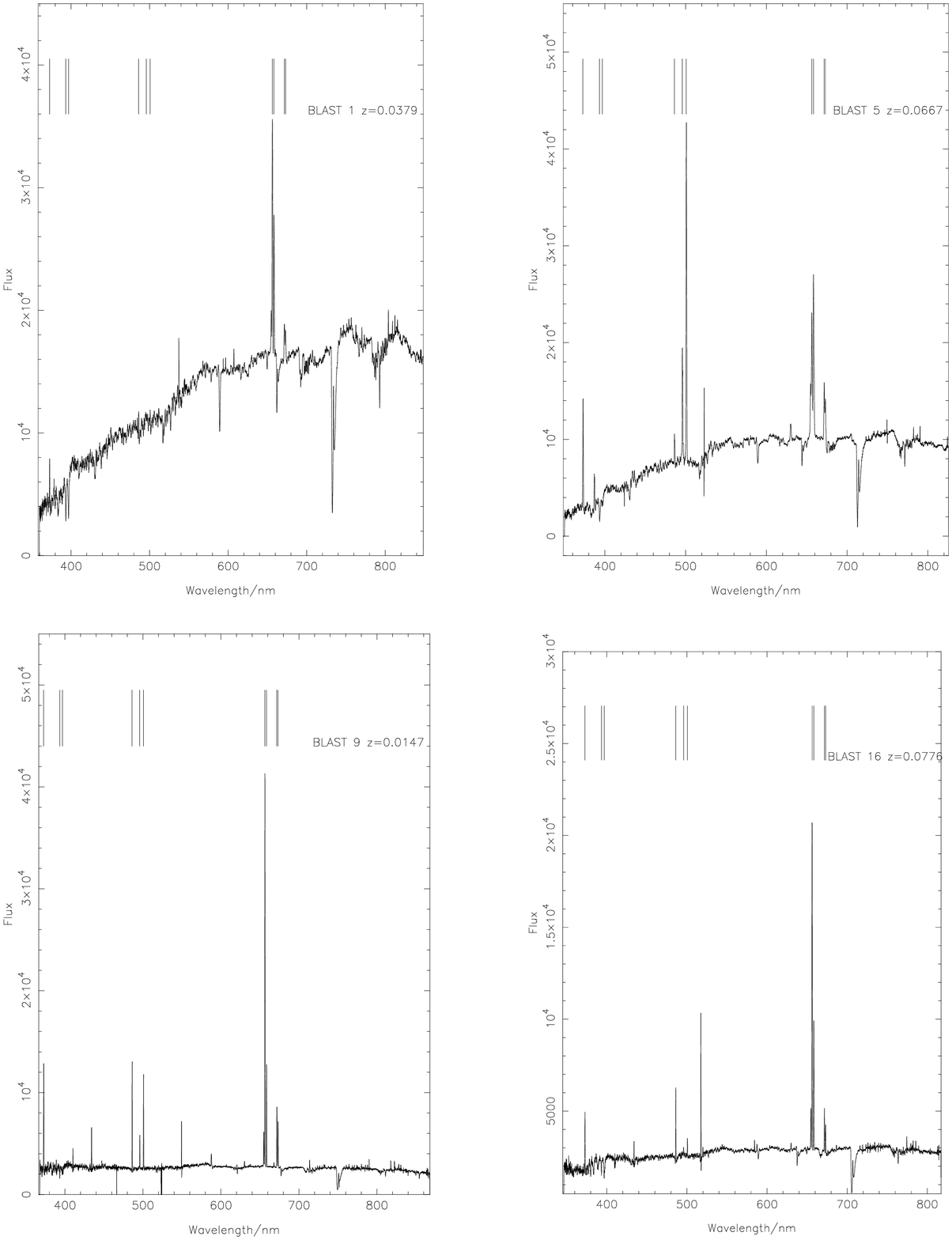}
\caption{Plots of the spectra of 1st, 6th, 11th... galaxies
listed in Table 2. The spectra are plotted in the rest frame of
each galaxy and the vertical lines show the positions
of the main features used to determine the redshifts.
From left to right, these are [OII] 372.7, the Calcium H and K lines,
H$\beta$, [OIII] 495.9 and 500.7, H$\alpha$, [NII] 658.3 and
[SII] 671.6 and 673.1.}
\end{figure}

\clearpage

\begin{figure}
\figurenum{2b}
\plotone{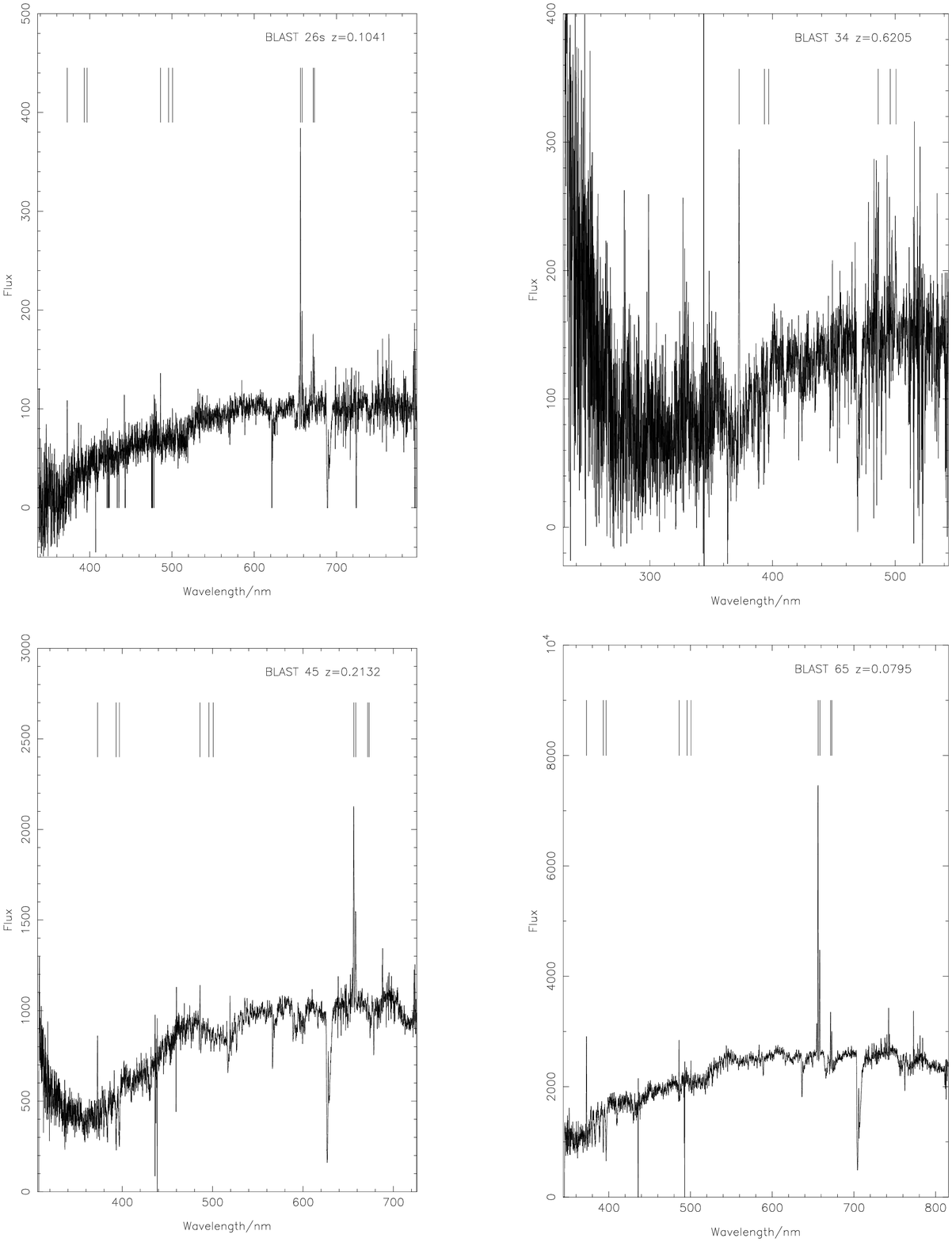}
\caption{Figure 2, continued}
\end{figure}
\clearpage

\begin{figure}
\figurenum{2c}
\plotone{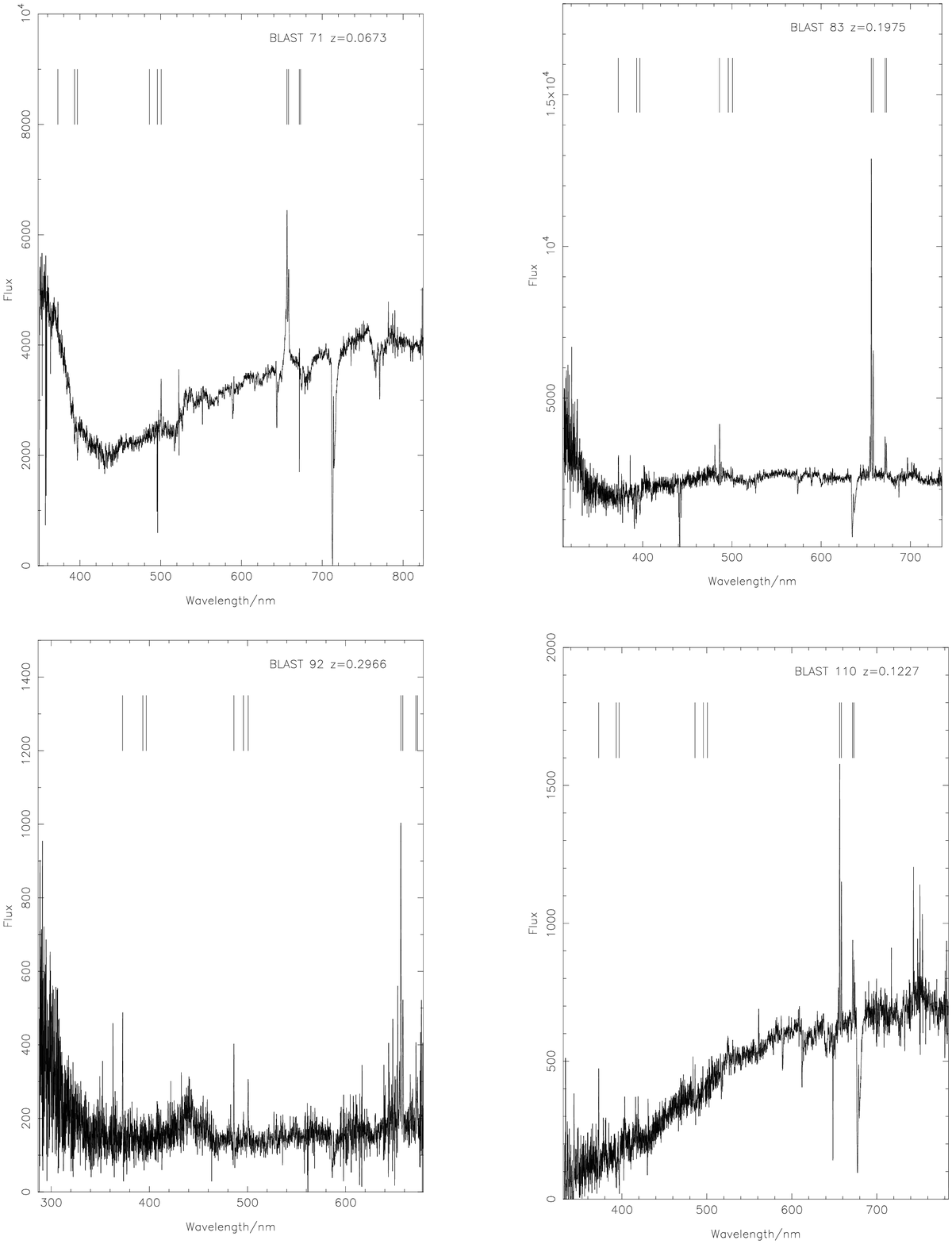}
\caption{Figure 2, continued}
\end{figure}
\clearpage
\begin{figure}
\figurenum{2d}
\plotone{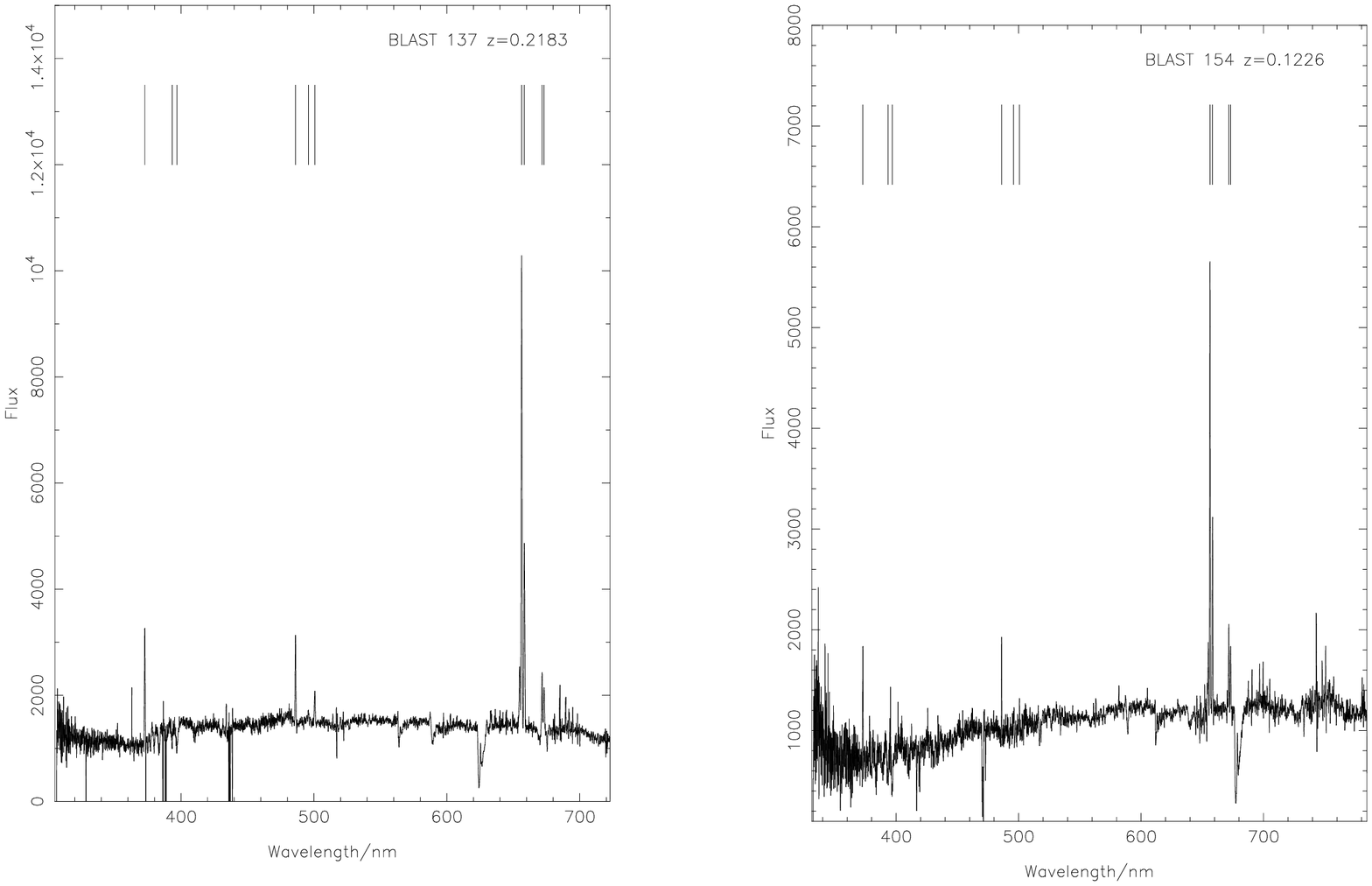}
\caption{Figure 2, continued}
\end{figure}

\clearpage

\clearpage

\LongTables
\begin{landscape}
\begin{deluxetable}{cccrrrrrrrrcrl}
\tabletypesize{\scriptsize}
\tablecaption{Secondary radio and 24-$\mu$m counterparts to 5$\sigma$ BLAST sources}
\tablewidth{0pt}
\tablehead{
\colhead{BLAST ID} & \colhead{Name} & \colhead{$\alpha$(radio)} & \colhead{$\delta$ (radio)} & 
\colhead{$f_r$} & \colhead{$P_r$} & \colhead{$d_r$} & \colhead{$\alpha$(24 $\mu$m)} & \colhead{$\delta$ (24 $\mu$m)}
& \colhead{$f_{24}$} & \colhead{$P_{24}$} & \colhead{$d_{24}$} & \colhead{$z_{RR}$} & \colhead{$z_{17}$} \\
}
\startdata
 2 & BLAST J032956-284631 & - & - & - & - & - &  52.48662 & -28.77239 &   9.936 &   0.00156 &     11.19 & - & -\\
  3 & BLAST J032741-282325 & - & - & - & - & - &  51.91867 & -28.39591 &   3.206 &   0.02433 &     22.36 &   0.132 & -\\
  4 & BLAST J033235-275530 &  53.14669 & -27.92555 &   0.05 &   0.01103 &   2.14 & - & - & - & - &    - &   0.062 &   0.038\\
  6 & BLAST J033229-274415 &  53.12498 & -27.73486 &   0.22 &   0.03452 &  10.75 &  53.12493 & -27.73467 &   4.620 &   0.00781 &     11.29 &   0.042 &   0.086\\
  8 & BLAST J033548-274920 & - & - & - & - & - &  53.95543 & -27.82605 &   3.666 &   0.01048 &     15.04 & - & -\\
  9 & BLAST J032916-273919 & - & - & - & - & - &  52.32330 & -27.65115 &   2.800 &   0.02386 &     19.41 &   0.143 & -\\
 12 & BLAST J032907-284121 &  52.28493 & -28.69235 &   1.70 &   0.00937 &  14.78 & - & - & - & - &    - & - & -\\
 16 & BLAST J033059-280955 & - & - & - & - & - &  52.74511 & -28.16186 &   0.900 &   0.05783 &     13.75 &   0.652 & -\\
 18 & BLAST J033123-275707 & - & - & - & - & - &  52.85186 & -27.95336 &   0.633 &   0.04198 &      8.07 &   0.419 &   0.495\\
 20 & BLAST J033340-273811 &  53.41863 & -27.64301 &   0.10 &   0.03013 &  23.92 & - & - & - & - &    - & - & -\\
 21 & BLAST J033152-281235 & - & - & - & - & - &  52.97013 & -28.20973 &   0.394 &   0.03855 &      4.94 &   0.288 & -\\
 24 & BLAST J033129-275720 &  52.87107 & -27.95562 &   0.05 &   0.04702 &   5.42 &  52.87106 & -27.95554 &   0.276 &   0.10290 &      5.48 &   1.070 &   0.767\\
 26 & BLAST J033246-275743 &  53.19105 & -27.96248 &   0.05 &   0.01507 &   2.56 & - & - & - & - &    - & - &   0.108\\
 30 & BLAST J033111-275820 &  52.79932 & -27.97172 &   0.10 &   0.00350 &   4.09 & - & - & - & - &    - &   0.493 & -\\
 32 & BLAST J033332-272900 & - & - & - & - & - &  53.38670 & -27.48273 &   0.336 &   0.08406 &      7.83 & - & -\\
 35 & BLAST J033217-275905 &  53.07121 & -27.98805 &   0.15 &   0.05232 &  11.51 &  53.07106 & -27.98794 &   2.050 &   0.02658 &     11.17 &   0.122 &   0.123\\
 37 & BLAST J032842-264107 & - & - & - & - & - &  52.18139 & -26.68347 &   2.356 &   0.01388 &     12.06 & - & -\\
 38 & BLAST J033216-280350 &  53.06599 & -28.06763 &   0.06 &   0.02123 &  14.03 &  53.06615 & -28.06751 &   0.288 &   0.16220 &     13.45 &   0.905 & -\\
 39 & BLAST J033106-274508 & - & - & - & - & - &  52.77509 & -27.75202 &   0.500 &   0.01179 &      2.90 & - & -\\
 44 & BLAST J033131-273235 &  52.87482 & -27.53938 &   0.10 &   0.03042 &  24.72 & - & - & - & - &    - & - & -\\
 45 & BLAST J033150-281126 & - & - & - & - & - &  52.96185 & -28.19172 &   0.327 &   0.03000 &      3.78 &   1.014 & -\\
 52 & BLAST J033214-281133 & - & - & - & - & - &  53.05822 & -28.19142 &   1.531 &   0.02405 &     12.03 &   0.271 & -\\
 53 & BLAST J033419-265319 & - & - & - & - & - &  53.58541 & -26.88726 &   1.598 &   0.03841 &     16.92 & - & -\\
 55 & BLAST J033129-275557 &  52.87536 & -27.93410 &   0.30 &   0.01339 &   6.32 &  52.87523 & -27.93395 &   0.991 &   0.01811 &      5.68 &   0.660 &   0.694\\
 57 & BLAST J033432-275140 & - & - & - & - & - &  53.63568 & -27.85617 &   2.096 &   0.03243 &     18.84 & - & -\\
 64 & BLAST J033240-280310 &  53.16388 & -28.05305 &   0.08 &   0.01483 &  10.75 &  53.16393 & -28.05327 &   0.323 &   0.12430 &     10.63 &   1.455 & -\\
 70 & BLAST J033111-284835 & - & - & - & - & - &  52.80250 & -28.81501 &   4.015 &   0.01644 &     21.09 &   0.132 & -\\
 73 & BLAST J033158-273519 &  52.99914 & -27.59106 &   0.05 &   0.02626 &  17.34 & - & - & - & - &    - & - &   1.062\\
 75 & BLAST J033115-273905 &  52.80825 & -27.65299 &   1.38 &   0.01244 &  16.58 & - & - & - & - &    - & - & -\\
 76 & BLAST J033328-273949 &  53.36866 & -27.66061 &   0.08 &   0.01799 &  12.70 & - & - & - & - &    - & - &   0.891\\
 77 & BLAST J033218-273138 &  53.07720 & -27.52958 &   0.14 &   0.00972 &   8.26 &  53.07726 & -27.52964 &   0.862 &   0.02986 &      8.45 & - & -0\\
 80 & BLAST J033156-284241 & - & - & - & - & - &  52.98177 & -28.70943 &   1.852 &   0.03864 &     19.18 &   0.349 & -\\
 83 & BLAST J033633-284223 & - & - & - & - & - &  54.14521 & -28.70720 &   3.046 &   0.01074 &     12.88 &   0.236 & -\\
 93 & BLAST J033408-273514 & - & - & - & - & - &  53.53581 & -27.59052 &   0.806 &   0.05419 &     11.94 & - & -\\
 95 & BLAST J033343-270918 & - & - & - & - & - &  53.43343 & -27.16089 &   1.279 &   0.08357 &     24.27 & - & -\\
 96 & BLAST J033336-272854 & - & - & - & - & - &  53.40747 & -27.47960 &   1.140 &   0.05479 &     16.13 & - & -\\
103 & BLAST J032707-270516 & - & - & - & - & - &  51.78596 & -27.08813 &   2.271 &   0.00872 &      8.83 & - & -\\
106 & BLAST J032704-280713 & - & - & - & - & - &  51.76532 & -28.12022 &   1.100 &   0.03176 &     10.83 & - & -\\
112 & BLAST J033241-273818 &  53.18000 & -27.63707 &  13.09 &   0.00803 &  18.53 & - & - & - & - &    - & - &   0.813\\
113 & BLAST J033347-273848 & - & - & - & - & - &  53.44796 & -27.64914 &   0.410 &   0.08373 &      8.87 & - & -\\
115 & BLAST J033128-280508 & - & - & - & - & - &  52.87123 & -28.08875 &   1.217 &   0.05477 &     17.13 &   0.047 & -\\
118 & BLAST J033238-273151 &  53.16146 & -27.53541 &   0.06 &   0.02455 &  16.43 &  53.16208 & -27.53529 &   0.676 &   0.10430 &     16.63 & - & -0\\
120 & BLAST J032703-282950 & - & - & - & - & - &  51.76067 & -28.49950 &   0.785 &   0.08079 &     15.54 & - & -\\
123 & BLAST J033112-265716 & - & - & - & - & - &  52.80191 & -26.95707 &   6.459 &   0.00189 &      8.72 & - & -\\
127 & BLAST J033224-291707 & - & - & - & - & - &  53.10748 & -29.28633 &   2.285 &   0.01683 &     13.19 &   0.038 & -\\
129 & BLAST J033225-284148 & - & - & - & - & - &  53.10985 & -28.69939 &   0.998 &   0.03757 &     11.14 &   0.528 & -\\
131 & BLAST J033200-273604 &  52.99850 & -27.60009 &   0.08 &   0.04090 &   6.45 & - & - & - & - &    - &   1.208 &   0.951\\
135 & BLAST J033134-282344 & - & - & - & - & - &  52.89462 & -28.39604 &   0.375 &   0.04696 &      5.46 &   0.294 & -\\
139 & BLAST J033626-270939 & - & - & - & - & - &  54.10915 & -27.15662 &   1.649 &   0.03315 &     15.76 & - & -\\
152 & BLAST J033648-271936 & - & - & - & - & - &  54.19638 & -27.32450 &   2.601 &   0.01668 &     14.66 & - & -\\
157 & BLAST J033609-280942 & - & - & - & - & - &  54.03825 & -28.16533 &   0.827 &   0.05743 &     12.70 & - & -\\
158 & BLAST J033307-281412 & - & - & - & - & - &  53.27776 & -28.23521 &   1.828 &   0.01069 &      8.46 &   0.871 & -\\
162 & BLAST J033154-274406 &  52.97686 & -27.73424 &   0.05 &   0.04552 &   5.15 &  52.97674 & -27.73408 &   0.105 &   0.27920 &      5.34 &   1.051 &   0.783\\
165 & BLAST J033605-293357 & - & - & - & - & - &  54.01957 & -29.56710 &   0.442 &   0.07899 &      9.06 &   0.330 & -\\
173 & BLAST J033132-281257 & - & - & - & - & - &  52.89003 & -28.21305 &   1.307 &   0.06320 &     20.19 & - & -\\
175 & BLAST J033619-272415 & - & - & - & - & - &  54.08273 & -27.40656 &   3.215 &   0.00652 &     10.05 & - & -\\
196 & BLAST J033211-280514 &  53.04457 & -28.08470 &   0.07 &   0.02547 &  17.38 &  53.04475 & -28.08507 &   0.459 &   0.15050 &     16.20 & - & -0\\
197 & BLAST J033335-273244 &  53.40302 & -27.54707 &   0.09 &   0.02585 &  18.38 &  53.40299 & -27.54736 &   0.515 &   0.15560 &     18.62 & - & -0\\
204 & BLAST J033336-274359 &  53.40413 & -27.73304 &   0.08 &   0.01217 &   9.15 & - & - & - & - &    - & - & -\\
205 & BLAST J032713-285101 & - & - & - & - & - &  51.80637 & -28.85232 &   0.534 &   0.05281 &      8.10 & - & -\\
212 & BLAST J033127-281027 & - & - & - & - & - &  52.87097 & -28.17383 &   1.302 &   0.03743 &     14.03 &   1.061 & -\\
238 & BLAST J032813-285930 & - & - & - & - & - &  52.05949 & -28.99373 &   0.390 &   0.08256 &      8.39 &   0.854 & -\\
246 & BLAST J033053-275704 & - & - & - & - & - &  52.72699 & -27.95084 &   0.627 &   0.06190 &     10.46 &   0.923 & -\\
253 & BLAST J032726-291936 & - & - & - & - & - &  51.85948 & -29.32255 &   1.937 &   0.02911 &     16.44 & - & -\\
257 & BLAST J032550-284919 & - & - & - & - & - &  51.46498 & -28.81971 &   0.565 &   0.09010 &     12.53 & - & -\\
265 & BLAST J033127-274430 &  52.86495 & -27.74436 &   0.20 &   0.03312 &   9.71 &  52.86470 & -27.74426 &   0.560 &   0.08980 &      9.63 & - & -0\\
304 & BLAST J033231-280437 &  53.13414 & -28.07417 &   0.13 &   0.01852 &  13.61 &  53.13503 & -28.07431 &   0.465 &   0.14060 &     15.32 & - & -0\\
320 & BLAST J032656-291615 & - & - & - & - & - &  51.73905 & -29.26493 &   1.459 &   0.06635 &     23.00 & - & -\\
339 & BLAST J033018-285124 & - & - & - & - & - &  52.58119 & -28.85509 &   0.550 &   0.08353 &     11.56 &   2.062 & -\\

\enddata
\tablecomments{Reading from the left, the columns are: BLAST identification
number, full name of source, radio coordinates, radio flux ($f_r$/mJy), probability of the radio
source being a chance alignment ($P_r$), radio radial offset ($d_r$/arcsec),
24-$\mu$m coordinates, 24-$\mu$m flux density ($f_{24}$/mJy), probability
of the 24-$\mu$m source being a chance alignment ($P_{24}$), 24-$\mu$m radial
offset ($d_{24}$/arcsec), photometric redshift from Rowan-Robison et al. (2008),
photometric redshift from Wolf et al. (2004).}
\end{deluxetable}


\clearpage

\LongTables
\begin{deluxetable}{lccccccccl}
\tabletypesize{\scriptsize}
\tablecaption{Spectroscopic Redshifts for the BLAST Counterparts}
\tablewidth{0pt}
\tablehead{
\colhead{BLAST Id} & \colhead{name of source} & \colhead{counterpart} & \colhead{$\alpha$} & \colhead{$\delta$}  &
\colhead{redshift} & \colhead{quality} & \colhead{H$\alpha$ EW} & \colhead{[NII] 658.3/H$\alpha$} & \colhead{Comment} \\
}
\startdata
1 & BLAST J032921-280803 &  1 & 52.33788 & -28.13343 & 0.0379 &   5& 12 & 0.58 & ... \\
  2 & BLAST J032956-284631 &  1 & 52.48567 & -28.77572 & 0.0370 &   5& 6 & 0.47 & ... \\
  3 & BLAST J032741-282325 &   1 & 51.92112 & -28.38893 & 0.0607 &   5& 17 & 0.78 & AGN \\
  4\tablenotemark{a} & BLAST J033235-275530 &   1 & 53.14622 & -27.92569 & 0.0376 &   5& 8 & 0.55 & ... \\
  4\tablenotemark{a} & BLAST J033235-275530 &   2 & 53.14669 & -27.92555 & 0.0379 &   4& 12 & 0.50 & ... \\
  5 & BLAST J033131-272842 &   1 & 52.88047 & -27.47975 & 0.0667 &   5& 17 & 1.29 & AGN \\
  6 & BLAST J033229-274415 &   1 & 53.12452 & -27.74028 & 0.0759 &   5& 40 & 0.43 & ... \\
  6 & BLAST J033229-274415 &   2 & 53.12498 & -27.73486 & 0.0755 &   5& 33 & 0.39 & ... \\
  7 & BLAST J033250-273420 &   1 & 53.20818 & -27.57581 & 0.2513 &   5& 30 & 0.45 & ... \\
  8 & BLAST J033548-274920 &   1 & 53.95480 & -27.82182 & 0.1675 &   4& ... & ... & ... \\
  9 & BLAST J032916-273919 &   1 & 52.31879 & -27.65604 & 0.0147 &   5& 85 & 0.27 & ... \\
 11 & BLAST J033424-274527 &   1 & 53.60244 & -27.75840 & 0.1245 &   5& 15 & 0.48 & ... \\
 12 & BLAST J032907-284121 &   1 & 52.28164 & -28.68818 & 0.0669 &   5& 9 & 0.59 & ... \\
 13 & BLAST J032950-285058 &   1 & 52.45617 & -28.84953 & 0.0761 &   5& 10 & 0.57 & ... \\
 15 & BLAST J033341-280742 &   1 & 53.42390 & -28.12707 & 0.3492 &   5& ... & ... & ... \\
 16 & BLAST J033059-280955 &   1 & 52.74791 & -28.16681 & 0.0776 &   5& 38 & 0.40 & ... \\
 19 & BLAST J033417-273927 &   1 & 53.57372 & -27.65888 & 0.1458 &   5& 25 & 0.49 & ... \\
 20 & BLAST J033340-273811 &   1 & 53.42222 & -27.63578 & 0.1015 &   5& 7 & 1.09 & AGN \\
 21 & BLAST J033152-281235 &   1 & 52.96505 & -28.20779 & 0.1809 &   5& 8 & 0.75 & AGN \\
 26\tablenotemark{a} & BLAST J033246-275743 &   1 & 53.19183 & -27.96262 & 0.1038 &   5&  16 & 0.36 & ... \\
 26\tablenotemark{a} & BLAST J033246-275743 &   2 & 53.19105 & -27.96248 & 0.1041 &   5& 16 & 0.37 & ... \\
 27 & BLAST J032956-281843 &   1 & 52.48742 & -28.31082 & 0.0595 &   5& 25 & 0.42 & ... \\
 29 & BLAST J032822-283205 &   1 & 52.09459 & -28.53271 & 0.0702 &   5& 17 & 0.43 & ... \\
 31 & BLAST J033414-274217 &   1 & 53.56034 & -27.70594 & 0.1027 &   5& 28 & 0.43 & ... \\
 32 & BLAST J033332-272900 &   1 & 53.38408 & -27.48811 & 0.1447 &   5& 43 & 0.47 & ... \\
 34 & BLAST J033149-274335 &   1 & 52.95710 & -27.72408 & 0.6205 &   5& ... & ... & ... \\
 35 & BLAST J033217-275905 &   2 & 53.07121 & -27.98805 & 0.1255 &   5& 6 & 0.72 & AGN \\
 38 & BLAST J033216-280350 &   1 & 53.06646 & -28.06318 & 0.5193 &   4& ... & ... & ... \\
 41 & BLAST J033430-271915 &   1 & 53.62771 & -27.32085 & 0.1033 &   5& 23 & 0.50 & ... \\
 43 & BLAST J033308-274809 &   1 & 53.29048 & -27.80045 & 0.1808 &   5& 33 & 0.31 & ... \\
 45 & BLAST J033150-281126 &   1 & 52.96213 & -28.18903 & 0.2132 &   5& 8 & 0.52 & ... \\
 49 & BLAST J033032-273527 &   1 & 52.63681 & -27.59523 & 0.1067 &   5& 21 & 0.44 & ... \\
 51 & BLAST J033046-275515 &   1 & 52.69279 & -27.92153 & 0.5245 &   5& ... & ... & ... \\
 55 & BLAST J033129-275557 &   2 & 52.87536 & -27.93410 & 0.6777 &   5& ... & ... & ... \\
 63 & BLAST J033316-275045 &   1 & 53.31882 & -27.84430 & 0.0874 &   5& 15 & 0.47 & ... \\
 65 & BLAST J033018-275500 &   1 & 52.57565 & -27.91658 & 0.0795 &   5& 11 & 0.42 & ... \\
 68 & BLAST J033146-275732 &   1 & 52.94418 & -27.95975 & 0.3645 &   5& ... & ... & ... \\
 69 & BLAST J033153-281036 &   1 & 52.97765 & -28.17654 & 0.2147 &   5& 33 & 0.44 & ... \\
 70 & BLAST J033111-284835 &   1 & 52.79586 & -28.80891 & 0.1089 &   5& 3 & 1.52 & AGN \\
 70 & BLAST J033111-284835 &   2 & 52.80220 & -28.81489 & 0.1093 &   5& 26 & 0.41 & ... \\
 71 & BLAST J033140-272937 &   1 & 52.91907 & -27.49373 & 0.0673 &   5& ... & ... & AGN (broad H$\alpha$)\\
 72 & BLAST J033120-273344 &   1 & 52.83482 & -27.56291 & 0.1950 &   5& 16 & 0.50 & ... \\
 75 & BLAST J033115-273905 &   1 & 52.81060 & -27.65189 & 0.3118 &   5& 8 & 0.65 & AGN? \\
 77 & BLAST J033218-273138 &   1 & 53.07986 & -27.52750 & 0.2272 &   5&  18 & 0.45 & ... \\
 80 & BLAST J033156-284241 &   2 & 52.98156 & -28.70936 & 0.4247 &   4& ... & ... & ... \\
 83 & BLAST J033633-284223 &   1 & 54.14325 & -28.70860 & 0.1975 &   5& 27 & 0.41 & ... \\
 84 & BLAST J033318-281436 &   1 & 53.32932 & -28.24242 & 0.1029 &   5& 13 & 0.51 & ... \\
 86 & BLAST J033447-283013 &   1 & 53.69997 & -28.50265 & 0.0414 &   5& 27 & 0.44 & ... \\
 88 & BLAST J033636-284115 &   1 & 54.15538 & -28.68720 & 0.0683 &   5& 37 & 0.43 & ... \\
 90 & BLAST J032818-274311 &   1 & 52.07532 & -27.71906 & 0.2484 &   5& 5 & 1.44 & AGN \\
92 & BLAST J033241-280557 &   1 & 53.17420 & -28.09792 & 0.2966 &   5& 22 & 0.44 & ... \\
 94 & BLAST J033351-274357 &   1 & 53.46999 & -27.72898 & 0.2250 &   5& 15 & 0.42 & ... \\
 95 & BLAST J033343-270918 &   1 & 53.42941 & -27.15325 & 0.0685 &   5& 5 & 0.65 & AGN? \\
 96 & BLAST J033336-272854 &   1 & 53.40470 & -27.48562 & 0.1449 &   5& 21 & 0.51 & ... \\
 97 & BLAST J033317-280220 &   1 & 53.31762 & -28.03985 & 0.3490 &   5& ... & ... & ... \\
110 & BLAST J033217-275054 &   1 & 53.07441 & -27.84972 & 0.1227 &   5& 9 & 0.55 & ... \\
122 & BLAST J033025-275014 &   1 & 52.60704 & -27.83831 & 0.1215 &   5& 36 & 0.36 & ... \\
126 & BLAST J033211-283251 &   1 & 53.05222 & -28.54655 & 0.6938 &   5& ... & ... & ... \\
129 & BLAST J033225-284148 &   1 & 53.11388 & -28.69935 & 0.1716 &   5& 31 & 0.45 & ... \\
135 & BLAST J033134-282344 &   1 & 52.89143 & -28.40074 & 0.2790 &   5& 73 & 0.40 & ... \\
137 & BLAST J032822-280809 &   1 & 52.08969 & -28.13662 & 0.2183 &   5& 50 & 0.43 & ... \\
139 & BLAST J033626-270939 &   1 & 54.10843 & -27.15991 & 0.2440 &   5& 48 & 0.66 & AGN? \\
143 & BLAST J033148-280958 &   1 & 52.95023 & -28.16929 & 0.3801 &   4& ... & ... & ... \\
149 & BLAST J033612-281046 &   1 & 54.05821 & -28.18282 & 0.1967 &   5& 16 & 0.79 & AGN \\
152 & BLAST J033648-271936 &   1 & 54.20436 & -27.32737 & 0.1458 &   5& 10 & 0.80 & AGN  \\
154 & BLAST J033541-285524 &   1 & 53.92151 & -28.92273 & 0.1226 &   5& 22 & 0.46 & ... \\
155 & BLAST J032929-284222 &   1 & 52.37314 & -28.70542 & 0.0703 &   5& 25 & 0.37 & ... \\
157 & BLAST J033609-280942 &   1 & 54.03806 & -28.16195 & 0.3159 &   5& 22 & 0.61 & ... \\
163 & BLAST J033114-273412 &   1 & 52.80927 & -27.57008 & 0.5336 &   5& ... & ... & ... \\
167 & BLAST J033247-274221 &   1 & 53.19950 & -27.70914 & 0.9805 &   4& ... & ... & ... \\
175 & BLAST J033619-272415 &   2 & 54.08242 & -27.40655 & 0.3431 &   3& ... & ... & ... \\
188 & BLAST J033111-275546 &   1 & 52.79504 & -27.93130 & 0.2815 &   5& 44 & 0.45 & ... \\
198 & BLAST J033215-273930 &   1 & 53.06759 & -27.65860 & 1.3236 &   4& ... & ... & quasar \\
207 & BLAST J033353-275555 &   1 & 53.47462 & -27.93015 & 1.9400 &   3& ... & ... & quasar? \\
212 & BLAST J033127-281027 &   2 & 52.87014 & -28.17357 & 0.8571 &   4& ... & ... & ... \\
221 & BLAST J033211-273729 &   1 & 53.04864 & -27.62401 & 1.5647 &   5& ... & ... & quasar \\
226 & BLAST J033723-274021 &   1 & 54.34529 & -27.67240 & 1.8017 &   5& ... & ... & quasar \\
259 & BLAST J033105-280634 &   1 & 52.77184 & -28.10405 & 0.1670 &   5& 25 & 0.61 & ...  \\
274 & BLAST J033053-275513 &   1 & 52.71999 & -27.91641 & 0.8950 &   4& ... & ... & ... \\
303 & BLAST J033121-275803 &   1 & 52.84258 & -27.96543 & 0.5297 &   5& ... & ... & AGN \\
329 & BLAST J033332-281348 &   1 & 53.39012 & -28.23444 & 1.3763 &   4& ... & ... & quasar\\
355 & BLAST J033117-272006 &   1 & 52.82410 & -27.33806 & 0.1064 &   5& 8 & 1.08 & AGN \\
\enddata
\tablenotetext{a}{The radio/24-$\mu$m counterparts for this source
fall within the same galaxy visible on the optical image (\S4.3), and so these redshifts
are effectively independent measurements for the same galaxy.}
\tablecomments{Reading from the left, the columns are:
the BLAST identification number; the full
name of the BLAST source; a number indicating whether
the counterpart is a primary or a secondary one (\S 2); the
position of the counterpart (in order of preference,
a radio, optical or IRAC 3.5-$\mu$m position); the redshift;
the quality of the redshift (A quality of 5 indicate the
redshift is certain; a quality of 4 indicates that the
redshift is almost certain---roughly a 95\% of being correct;
a quality of 3 indicates that the redshift is somewhat less
certain but still probably correct.); 
the equivalent width of the H$\alpha$ line in the rest frame; the
ratio of the flux in the [NII] 658.3 line to the flux in the
H$\alpha$ line;
a comment on whether there is any evidence from the line ratios
or the width of the lines that the galaxy contains an AGN (see \S 4.1 for details).}
\end{deluxetable}

\clearpage
\end{landscape}

\begin{deluxetable}{cccc}
\tabletypesize{\scriptsize}
\tablecaption{Spectroscopic Redshifts for the Other Targets}
\tablewidth{0pt}
\tablehead{
\colhead{$\alpha$} & \colhead{$\delta$} & \colhead{Redshift} & \colhead{quality} \\
}
\startdata
53.41404 & -28.29011 & 0.9851 &   3\\
53.04004 & -28.12106 & 0.9805 &   3\\
53.31208 & -28.32286 & 0.9298 &   3\\
53.70525 & -28.29303 & 0.1489 &   5\\
53.54129 & -28.29753 & 0.8305 &   3\\
53.74600 & -28.39853 & 0.3492 &   4\\
53.53217 & -28.35928 & 0.3084 &   5\\
53.65284 & -28.42636 & 0.3614 &   5\\
53.53867 & -28.40558 & 0.6967 &   5\\
53.56683 & -28.45158 & 0.3816 &   3\\
53.56208 & -28.43239 & 0.1031 &   5\\
53.60571 & -28.50286 & 0.7325 &   4\\
53.42112 & -28.40719 & 0.2890 &   5\\
53.24021 & -28.38847 & 0.2136 &   5\\
..... & ..... & ....& .... \\
\enddata
\tablecomments{Reading from the left, the columns are:
the position of the fibre on the sky; the redshfit; the quality of the redshift (A quality of 5 indicate the
redshift is certain; a quality of 4 indicates that the
redshift is almost certain---roughly a 95\% of being correct;
a quality of 3 indicates that the redshift is somewhat less
certain but still probably correct) }
\end{deluxetable}



\begin{deluxetable}{cccc}
\tabletypesize{\scriptsize}
\tablecaption{The 5$\sigma$ Samples at Each Wavelength}
\tablewidth{0pt}
\tablehead{
\colhead{Wavelength} & \colhead{Sources} & \colhead{Counterparts} & \colhead{Redshifts} \\
}
\startdata
250 $\mu$m & 115 & 94 & 82 (49) \\
350 $\mu$m & 89 & 62 & 48 (27) \\
500 $\mu$m & 107 & 52 & 39 (12) \\
\enddata
\tablecomments{Reading from the left, the columns are:
the wavelength of the sample; the number of sources
detected at $>$5$\sigma$ at this wavelength and that
are in an area covered by deep optical images; the number
of these sources with radio and 24 $\mu$m counterparts or
both; the number of counterparts with either photometric or
spectroscopic redshifts (the number of spectroscopic
redshifts is in brackets).}
\end{deluxetable}




\end{document}